\documentclass[iop,twocolumn,twocolappendix,numberedappendix]{emulateapj}
\usepackage{epsfig}
\usepackage{amsmath}
\usepackage{amssymb}
\usepackage{natbib}
\usepackage{hyperref}
\usepackage{graphicx}
\usepackage{times}
  \usepackage[flushleft]{threeparttable}

\newcommand{\coone}{\mbox{CO\,(1--0)}}
\newcommand{\cotwo}{\mbox{CO\,(2--1)}}
\begin{document}
\defcitealias{Querejeta2015}{Q15}

\title{
The organization of cloud-scale gas density structure: high resolution CO vs. 3.6~\boldmath{$\mu$}m brightness contrasts in nearby galaxies} 
\author{Sharon E. Meidt\altaffilmark{1}}
\altaffiltext{1}{Sterrenkundig Observatorium, Universiteit Gent, Krijgslaan 281 S9, B-9000 Gent, Belgium}

\author{Adam K. Leroy\altaffilmark{2}}
\altaffiltext{2}{Department of Astronomy, The Ohio State University, 140 W. 18th Ave., Columbus, OH 43210, USA}

\author{Miguel Querejeta\altaffilmark{3}}
\altaffiltext{3}{Observatorio Astron\'{o}mico Nacional - (IGN), Observatorio de Madrid Alfonso XII, 3, 28014 - Madrid, Spain}

\author{Eva Schinnerer\altaffilmark{4}}
\altaffiltext{}{Max-Planck-Institut f\"ur Astronomie, K\"{o}nigstuhl 17, D-69117 Heidelberg, Germany}

\author{Jiayi Sun\altaffilmark{2}}

\author{Arjen van der Wel\altaffilmark{1}}

\author{Eric Emsellem\altaffilmark{5,6}}
\altaffiltext{5}{European Southern Observatory, Karl-Schwarzschild-Stra{\ss}e 2, D-85748 Garching, Germany}
\altaffiltext{6}{Univ Lyon, Univ Lyon1, ENS de Lyon, CNRS, Centre de Recherche Astrophysique de Lyon UMR5574, F-69230 Saint-Genis-Laval France}

\author{Jonathan Henshaw\altaffilmark{4}}

\author{Annie Hughes\altaffilmark{7,8}}
\altaffiltext{7}{CNRS, IRAP, 9 Av. du Colonel Roche, BP 44346, F-31028 Toulouse cedex 4, France}
\altaffiltext{8}{Universit\'{e} de Toulouse, UPS-OMP, IRAP, F-31028 Toulouse cedex 4, France}

\author{J.~M.~Diederik Kruijssen\altaffilmark{9}}
\altaffiltext{9}{Astronomisches Rechen-Institut, Zentrum f\" ur Astronomie der Universit\"at Heidelberg, Department of M\"onchhofstra\ss e 12-14, D-69120 Heidelberg, Germany}

\author{Erik Rosolowsky\altaffilmark{10}}
\altaffiltext{10}{4-183 CCIS, University of Alberta, Edmonton, Alberta, Canada}

\author{Andreas Schruba\altaffilmark{11}}
\altaffiltext{11}{Max-Planck-Institut f\"ur extraterrestrische Physik, Giessenbachstra\ss e 1, D-85748 Garching, Germany}

\author{Ashley Barnes\altaffilmark{12}}
\altaffiltext{12}{Argelander-Institut für Astronomie, Universität Bonn, Auf dem Hügel 71, D-53121 Bonn, Germany}

\author{Frank Bigiel\altaffilmark{12}}

\author{Guillermo A. Blanc\altaffilmark{13,14,15}}
\altaffiltext{13}{Departamento de Astronom\'{i}a, Universidad de Chile, Casilla 36-D, Santiago, Chile}
\altaffiltext{14}{Centro de Astrof\'{i}sica y Tecnolog\'{i}as Afines (CATA), Camino del Observatorio 1515, Las Condes, Santiago, Chile}
\altaffiltext{15}{Visiting Astronomer, Observatories of the Carnegie Institution for Science, 813 Santa Barbara St, Pasadena, CA, 91101, USA}
\
\author{Melanie Chevance\altaffilmark{9}}
\altaffiltext{9}{Astronomisches Rechen-Institut, Zentrum f\" ur Astronomie der Universit\"at Heidelberg, Department of M\"onchhofstra\ss e 12-14, D-69120 Heidelberg, Germany}

\author{Yixian Cao\altaffilmark{16}}
\altaffiltext{18}{Aix Marseille Université, CNRS, LAM (Laboratoire d’Astrophysique de Marseille), F-13388 Marseille, France}

\author{Daniel A. Dale\altaffilmark{17}}
\altaffiltext{17}{Physics \& Astronomy, University of Wyoming, Laramie, WY 82071}

\author{Christopher Faesi\altaffilmark{4,18}}
\altaffiltext{18}{Department of Astronomy, University of Massachusetts - Amherst, 710 N. Pleasant St., Amherst, MA 01003}

\author{Simon C.~O.~Glover\altaffilmark{19}}
\altaffiltext{19}{Universit\"{a}t Heidelberg, Zentrum f\"{u}r Astronomie, Albert-Ueberle-Str. 2, D-69120 Heidelberg, Germany}

\author{Kathryn Grasha\altaffilmark{20}}
\altaffiltext{20}{Research School of Astronomy and Astrophysics, Australian National University, Canberra, ACT 2611, Australia}

\author{Brent Groves\altaffilmark{20,21}}
\altaffiltext{21}{International Centre for Radio Astronomy Research, University of Western Australia, 7 Fairway, Crawley, 6009, WA, Australia}

\author{Cinthya Herrera\altaffilmark{22}}
\altaffiltext{}{Institut de Radioastronomie Millim\'etrique, 300 Rue de la Piscine, F-38406 Saint Martin d'H\`eres, France}

\author{Ralf S.\ Klessen\altaffilmark{19,23}}
\altaffiltext{23}{Universit\"{a}t Heidelberg, Interdisziplin\"{a}res Zentrum f\"{u}r Wissenschaftliches Rechnen, INF 205, D-69120 Heidelberg, Germany}

\author{Kathryn Kreckel\altaffilmark{9}}

\author{Daizhong Liu\altaffilmark{4}}

\author{Hsi-An Pan\altaffilmark{4}}

\author{Jerome Pety\altaffilmark{24}}
\altaffiltext{}{Institut de Radioastronomie Millim\'etrique, 300 Rue de la Piscine, F-38406 Saint Martin d'H\`eres, France}

\author{Toshiki Saito\altaffilmark{4}}

\author{Antonio Usero\altaffilmark{3}}

\author{Elizabeth Watkins\altaffilmark{9}}

\author{Thomas G. Williams\altaffilmark{4}}

\begin{abstract}
In this paper we examine the factors that shape the distribution of molecular gas surface densities on the 150~pc scale across 67 morphologically diverse star-forming galaxies in the PHANGS-ALMA \cotwo\ survey.  Dividing each galaxy into radial bins, we measure molecular gas surface density contrasts, defined here as the ratio between a fixed high percentile of the CO distribution and a fixed reference level in each bin.This reference level captures the level of the faint CO floor that extends between bright filamentary features, while the intensity level of the higher percentile probes the structures visually associated with bright, dense ISM features like spiral arms, bars, and filaments.  We compare these contrasts to matched percentile-based measurements of the 3.6~$\mu$m emission measured using \textit{Spitzer}/IRAC imaging, which trace the underlying stellar mass density. We find that the logarithms of CO contrasts on 150~pc scales are 3-4 times larger than, and positively correlated with, the logarithms of 3.6~$\mu$m contrasts probing smooth non-axisymmetric stellar bar and spiral structures.  The correlation appears steeper than linear, consistent with the compression of gas as it flows supersonically in response to large-scale stellar structures, even in the presence of weak or flocculent spiral arms.  Stellar dynamical features appear to play an important role in setting the cloud-scale gas density in our galaxies, with gas self-gravity perhaps playing a weaker role in setting the 150~pc-scale distribution of gas densities.
\end{abstract}

\section{Introduction\label{sec:intro}}
\setcounter{footnote}{0}
Dynamical features in galaxy disks like bars and spiral arms play an important role in organizing the cold neutral medium from which a galaxy forms new stars.   This is evident in the prominent chains of massive star formation that string out along spiral arms and punctuate large-scale bars \citep[e.g.,][]{elmelm83}, and more directly visible in the morphology of atomic and molecular gas tracers on $\gtrsim\!500$~pc scales in resolved surveys of nearby galaxies \citep{helfer2003, Walter2008, Leroy2009, donovan12}).  

Stellar features not only sweep up and organize the large-scale distribution of cold gas in galaxies, they also influence the dynamical state of the gas on the scale of individual molecular clouds, thus affecting the ability of a galaxy's molecular cloud population to collapse and form stars 
\citep{hunter98,dobbsbonnel07,meidt13,jog2014,semenov17,gensior20,jeffreson20,Meidt2020}.  Through their influence on where high-mass star formation occurs in galaxies, large-scale stellar dynamical features also determine the sites of stellar feedback and in this manner they further regulate the ability of gas to form stars \citep{ostriker,hopkins,semenov18,kko2020}.

To understand the role of large-scale dynamical features in the star formation process, we need to examine how gas is organized and structured at `cloud scales,' a term that has been used to refer to the ${\sim}100$~pc scales where an individual resolution element corresponds to about the size of a massive giant molecular cloud or cloud complex \citep[e.g.,][]{Leroy2016}.  
Recent work showing that cloud-scale properties (including dynamical state) depend on galactic environment (\citealt{hughesII}; \citealt{colombo2014a}; \citealt{Leroy2017}; \citealt{Sun2018,Sun2020b}; \citealt{Schruba2019}; \citealt{Rosolowsky2021}) seems to indicate that bars and spirals  
have an influence on gas structure and organization that persists down to the cloud scale.  Such an influence is consistent with indications that integrated (time-averaged) depletion times (or inverse star formation efficiencies; SFEs) are not universal but also vary strongly with galactic environment (\citealt{Leroy2017}; \citealt{utomo17, utomo18}; \citealt{tomicic}; D.~Utomo et al.\ in prep.; but see \citealt{foyle10,moore,eden,tress,urquhart} for results indicating similar molecular gas SFEs in spiral arm and interarm or `flocculent' regions).

Measuring the range of spatial scales over which the background distribution of gas, stars and dark matter is the primary agent that organizes the cold gas reservoir in galaxies is also important for understanding the emergence of near ubiquitous exponential molecular gas surface density radial profiles \citep[e.g.,][]{Leroy2008} from the small-scale molecular gas distributions that appear to be constituted by clumpy, seemingly decoupled, discrete objects \citep[see also][]{elmegreen05}.  Recent work has emphasized the role of the turbulent pressure in the local ambient medium and large-scale gravitational field in establishing cloud-scale surface densities \citep{Sun2020a}, which themselves influence the highest densities that can be built within clouds (\citealt{gallagher}; and see \citealt{Meidt2016} and \citealt{burkmocz19}).  This connection is borne out in the kinematics of gas observed on ${\sim}400$~pc scales down to $0.1$~pc scales, which reveal a clear link between gas motions on roughly the disc scale height (or the region separation length; \citealt{Kruijssen2019b,Chevance2020b}) and the sonic scale \citep{henshaw20}. Together, these recent results suggest that large-scale galactic environment plays a much more prominent role on the cloud scale (and below) than previously thought.  

\begin{figure*}[t]
\begin{center}
\begin{tabular}{cc}
\hspace*{-.5cm}\includegraphics[width=.485\linewidth]{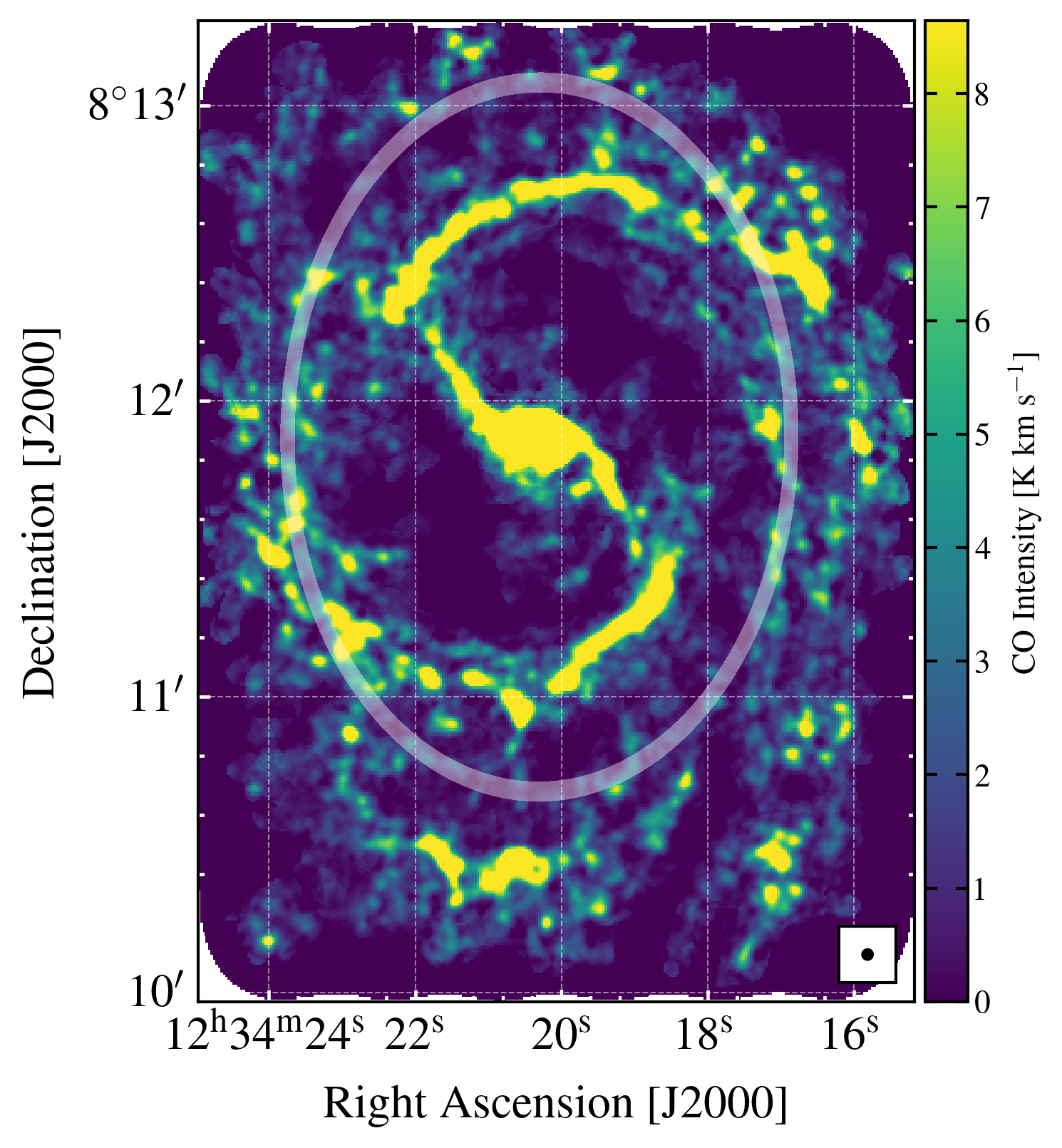}&
\includegraphics[width=.485\linewidth]{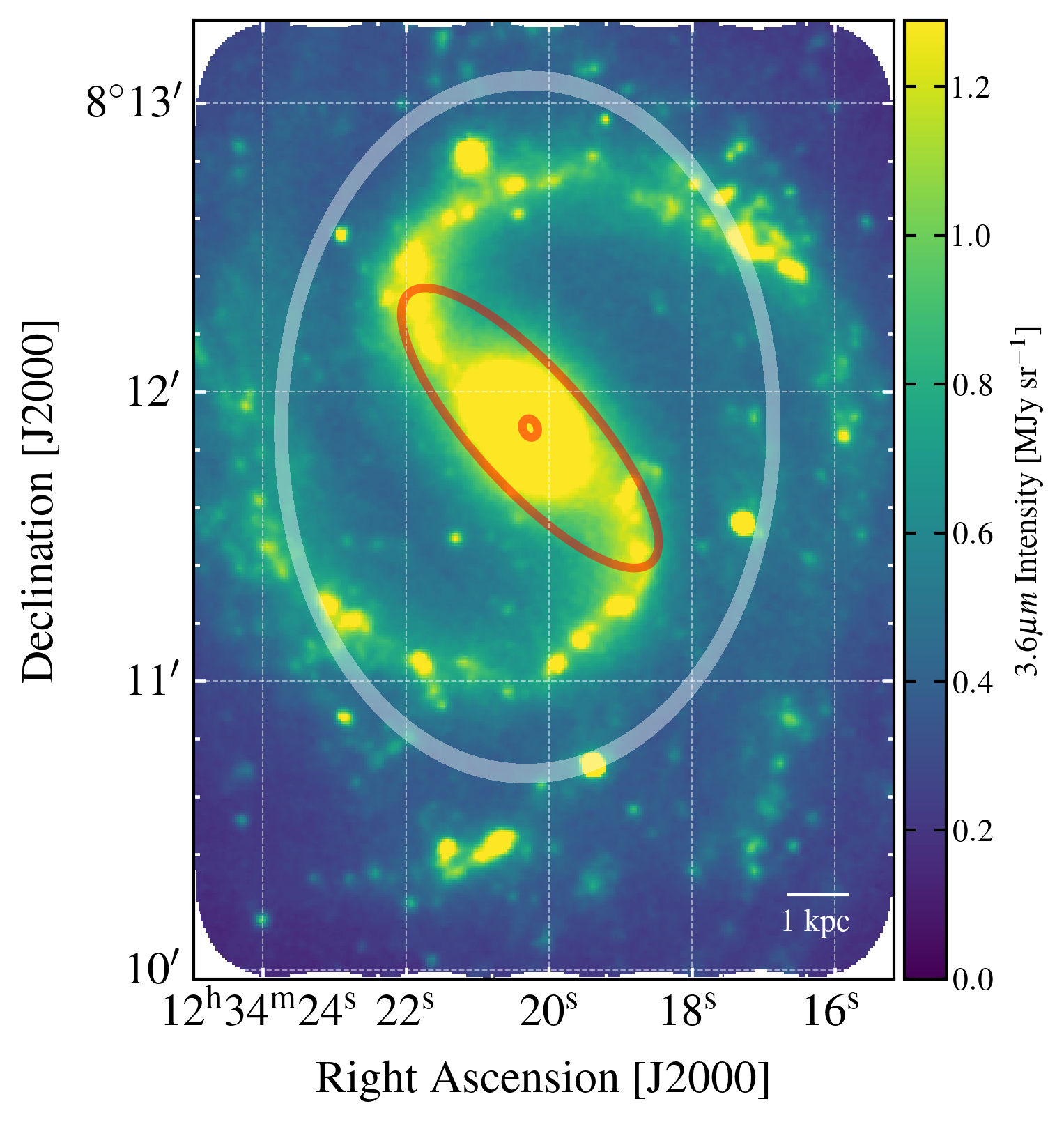}
\end{tabular}
\end{center}
\caption{Example (left) CO and (right) \textit{Spitzer}/IRAC 3.6~$\mu$m imaging distribution in one target, the spiral galaxy NGC~4535. The CO image shows the PHANGS-ALMA `broad' integrated intensity map. The 3.6~$\mu$m image shows the S$^4$G image. Both images have the same astrometry and are displayed on a linear stretch from 0 to the 95th percentile of the map, so if the maps were identical they would appear the same. Red ellipses in the right panel mark the central stellar overdensity and bar definitions used in this work \citep[see][]{Querejeta} The light gray ellipse in both panels marks the annulus at $R_{\rm gal}=5.5$ kpc shown in Figure~\ref{fig:exdists}.}
\label{fig:prettymap}
\end{figure*}

Another clear sign that the stellar potential exerts an influence on the cold gas is discernible in the azimuthal structure of CO emission. In high resolution, cloud-scale maps of CO, such as that shown in Figure~\ref{fig:prettymap}, CO emission tends to show a striking qualitative correspondence to structures in the stellar disk. So far, most of the studies listed above have emphasized the properties of detected molecular clouds or cloud-scale CO emission, placing less emphasis on the overall structure of the CO distribution, including non-detections and weak detections. But to the eye, this clear correspondence between sharp, well-defined CO emission and enhancements in the stellar potential well is one of the most striking results from the recent explosion of `cloud-scale' ALMA CO mapping of nearby galaxies. The quantitative link between the azimuthal CO structure at ${\sim}100$~pc resolution and stellar structure remains almost unexplored.

In this paper, we examine the factors that shape the azimuthal distribution of molecular gas throughout nearby galaxies surveyed by PHANGS-ALMA.  We use the contrast in gas surface density present at a fixed $150$~pc scale within galactic annuli (sampling from 0 to 2$\pi$ in azimuth at a given galactocentric radius), to pick out the presence of large-scale non-axisymmetry in the CO emission that we use to trace the bulk of the molecular gas.  We then compare our CO contrasts to ``reference'' contrasts measured from near-infrared (near-IR) 3.6~$\mu$m emission that probes the prominence of non-axisymmetric stellar structures like bars and spiral arms \citep[e.g.,][]{elm2011,bittner17}. In this way we quantify how gas near the cloud scale is organized by large-scale dynamical features together with self-gravity.

\section{The sample and the data}\label{sec:data}

We carry out statistical characterization of the amplitude of azimuthal variations in CO brightness and 3.6$\mu$m brightness. On their own, these measurements capture the strength of non-axisymmetric structures like spiral arms and stellar bars. The range in 3.6~$\mu$m brightness present at a given radial location in nearby galaxies has been used before to probe the prominence such features \citep[e.g.,][]{elm2011,bittner17}. In this paper we use the stellar contrast measured from the 3.6$\mu$m images as a baseline. We contrast the CO brightness variations against these $3.6\mu$m measurements with the aim of revealing the degree to which the organization of the cold gas is either passively or actively arranged by stellar features. Because our goal is to conduct a statistical analysis across a large part of the PHANGS--ALMA sample, we do not focus only on grand-design spirals or, e.g.,  spiral arm and interarm regions identified by eye. However, as we discuss below, our results do appear broadly consistent with those approaches.

\subsection{150~pc scale CO intensities}
The PHANGS-ALMA survey has mapped the distribution of \cotwo\ emission, tracing molecular gas, across 90 nearby main-sequence galaxies at a typical resolution of $1.2\arcsec$.  The sample selection and survey are described in \citet{Leroy2021b} and the data processing in \citet{Leroy2021a}.  The CO data have FWHM beam sizes corresponding to 100$_{-35}^{+31}$ pc scales after adopting the distances to our galaxies published by \citet{Anand21}.  

In this paper, we focus on the 67 galaxies studied by \citet{Lang20}, which represent a subset of PHANGS-ALMA with clear disk structure, spatially extended CO emission, and high quality near-IR data, all of which are key elements in our analysis (see below).  
We use the 150~pc resolution `broad mask' integrated \cotwo\ intensity maps from PHANGS-ALMA internal data release v3p4. These maps have high completeness, on average recovering almost all of the total CO flux (98\% with a 5-95th percentile range of 73\%-100\%) measured for our targets by direct integration of the cube (see \citealt{Leroy2021a} for more details). Using these maps, we expect to trace all of the CO emission at a fixed spatial resolution across our whole sample. Figure~\ref{fig:prettymap} shows examples of a CO map and a 3.6~$\mu$m map for one target galaxy.  

We are almost exclusively interested in surface density contrasts at fixed radius. Therefore we do not apply a CO-to-H$_2$ conversion factor $\alpha_{\rm CO}$, and work only with CO intensity. We expect most variations in $\alpha_{\rm CO}$ to be global or radial in nature, tracing metallicity gradients or large scale structure \citep[e.g., see][]{Sun2020a}. Large azimuthal variations at fixed galactocentric radius are not expected.  

\subsection{\texorpdfstring{3.6~$\mu$m}{3.6 micron} maps tracing the old stellar distribution}
To trace the structure of the underlying stellar disk, we use near-IR 3.6~$\mu$m \textit{Spitzer}/IRAC maps at ${\sim}1.7\arcsec$ resolution obtained in part by the S$^4$G survey \citep{Sheth2010} and otherwise as described in \cite{Querejeta}. These maps are dominated by the light from old stars and probe $36{-}215$~pc scales at the distances of our targets. For 57 of these IRAC images, the physical resolution is finer than 150~pc scale accessed by our CO data and so we construct a new set of maps convolved to 150~pc and regridded to the PHANGS-ALMA pixel grid.  For the remaining 10 galaxies in our sample, the 3.6~$\mu$m resolution is coarser than 150~pc. Since the distribution of old stars is expected to be intrinsically smooth, we assume that the light in these IRAC 3.6~$\mu$m images is distributed smoothly beneath the native resolution of the IRAC data and simply reproject this set of IRAC images onto the same pixel grid as the corresponding PHANGS-ALMA CO map.  This compromise should have little, if any, effect on the measured contrasts, given that further convolving the set of 150~pc-scale 3.6~$\mu$m images to the lowest map resolution (215 pc) yields no substantial difference in the measured contrasts (3\% scatter, with no systematic offset).  We conclude that retaining all galaxies in the sample is more important than requiring the CO and 3.6~$\mu$m maps to be strictly resolution matched, in order to investigate the importance of stellar structure across a representative sample of nearby galaxies.

Assuming that the mass-to-light ratio $\Upsilon_{3.6 \mu {\rm m}}$ does not vary substantially with azimuth in our targets, our measurement of a 3.6~$\mu$m intensity contrast within an annulus corresponds to a contrast in stellar mass surface density. For a reasonable spread in stellar age and metallicity, variations of $\Upsilon_{3.6 \mu {\rm m}}$ are expected to be small  \citep{mcgaughschom,meidt14,Leroy2019}. 
Highly localized contamination from non-stellar emission \citep[][hereafter \citetalias{Querejeta2015}]{Meidt12,Querejeta2015} is avoided using the strategy described in \S\ref{sec:contrasts}, verified by comparing with 3.6~$\mu$m stellar mass maps constructed by \citetalias{Querejeta2015}, where available.  These maps have resolutions double the native IRAC resolution, and we match them to the CO maps using the same scheme that we applied to the original 3.6~$\mu$m images.   
To remain as close as possible to the data, we adopt the original maps as our fiducial stellar tracer.

\subsection{Environment definitions}
We use the PHANGS 2D environmental masks \citep{Querejeta} to sort lines-of-sight by galactic environment.  These masks demarcate regions dominated by different stellar morphological features, in large part leveraging the 3.6~$\mu$m structural decompositions performed by \cite{Salo2015} and catalogs from \cite{Herrera2015}.  Using these masks, we separate regions dominated by bars from the rest of the disk (including spirals) and omit central regions (including bulges, lenses and rings; totalling 10\% of the area surveyed in CO) from our analysis. 
We also sort spirals according to arm class (grand design, multi-arm, flocculent) from \citet{buta15}.  For galaxies not in S$^4$G, arm classification follows the \citet{buta15} scheme.  
\subsection{Radial binning}
All maps are sampled in a series of radial bins oriented in the sky-plane at the inclination and position angle determined from PHANGS-ALMA CO kinematics by \cite{Lang20}.  These elliptical annuli have 150~pc width along the minor axis (i.e.\ the fiducial physical resolution set by the CO maps) and are spaced by an equal amount.  

\begin{figure}[t]
\begin{center}
\begin{tabular}{c}
\includegraphics[width=0.985\linewidth]{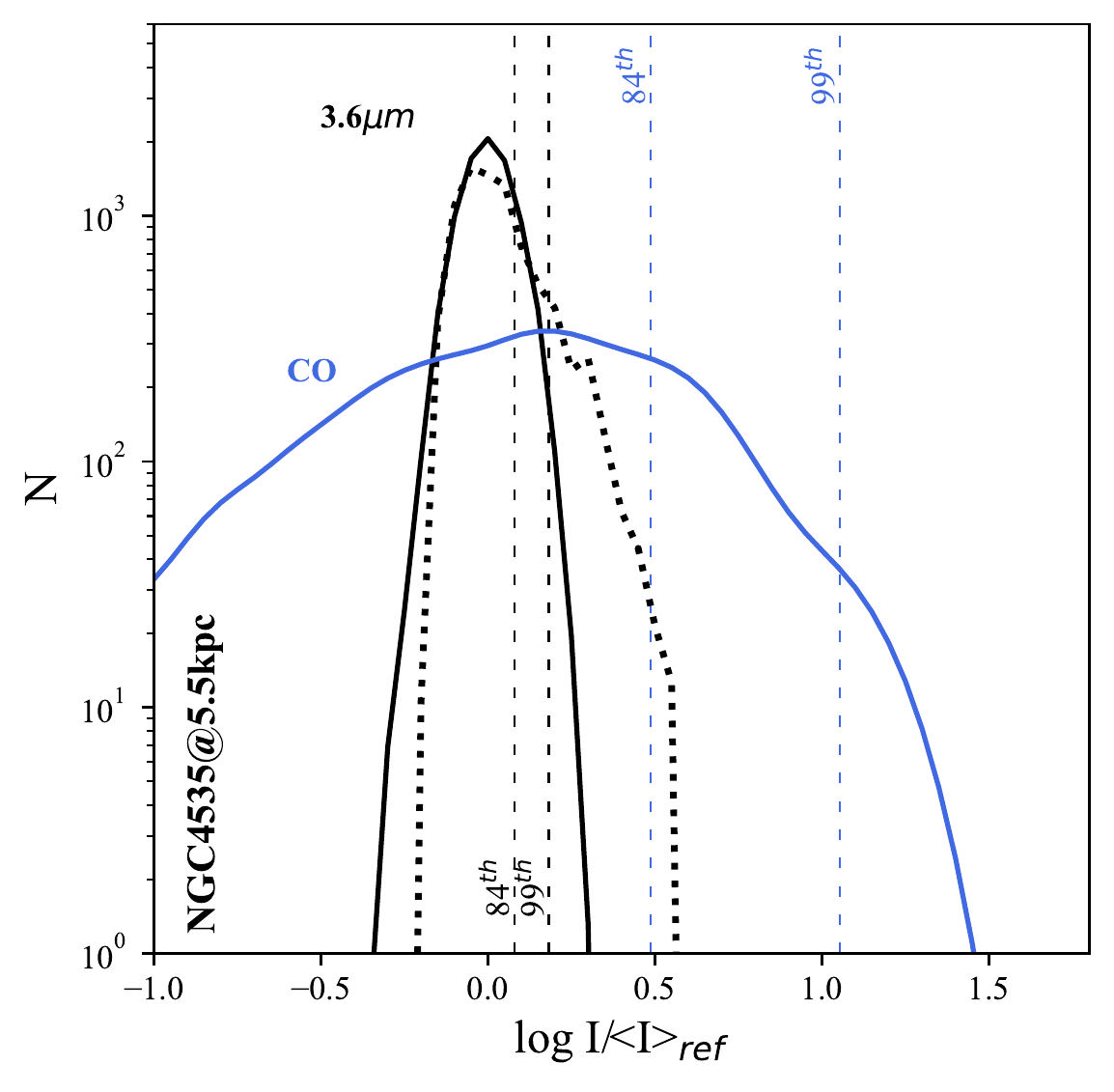}
\end{tabular}
\end{center}
\caption{Example CO (blue) and 3.6~$\mu$m (black) brightness distributions in an annulus at $R_{\rm gal}=5.5$~kpc $\pm150$~pc (marked in Figure~\ref{fig:prettymap}) in the spiral galaxy NGC~4535.  Distributions are normalized to the reference level ${\langle I\rangle}_{\rm ref}$ measured from each distribution as the mean brightness below the 84th percentile (see text for details). The solid black line shows the baseline log-normal 3.6~$\mu$m model used for contrast measurements, which we adopt to avoid high-brightness contamination from dust heated by star formation.  The observed 3.6~$\mu$m distribution is shown as a black dotted line.  Dashed vertical lines show the locations of the percentiles that we use to measure contrasts.}
\label{fig:exdists}
\end{figure}

\section{Characterizing brightness distributions with contrasts}\label{sec:contrasts}
\subsection{Typical shapes of CO and \texorpdfstring{3.6~$\mu$m}{3.6 micron} brightness distributions}
Representative CO and 3.6~$\mu$m intensity distributions for one annulus are shown in Figure~\ref{fig:exdists}. In the figure, we plot the normalized intensity distribution for pixels within an annulus centred at $R_{\rm gal}=5.5$~kpc in NGC~4535, a galaxy featuring a strong two-armed spiral pattern.  
 The wide CO distribution in this example resembles a log-normal, but across our sample, the CO distributions for some radial bins in some galaxies also exhibit a high-brightness tail (see \citealt{hughesI}).  
The stellar distribution shown in Figure~\ref{fig:exdists} is considerably narrower in width and also roughly log-normal, with the exception of a modest high brightness excess (described in \S\ref{sec:pctlprobes}). The distributions capture the visual impression from the maps themselves: the 3.6~$\mu$m distribution is much smoother (less strongly varying) than the CO map. The CO map presents a sharpened view of the common global spiral morphology shared by the two tracers (see Figure~\ref{fig:prettymap}). 

\subsection{CO and \texorpdfstring{3.6~$\mu$m}{3.6 micron} brightness contrasts}

\subsubsection{Definition of the contrast}
\label{sec:thecontrastdef}

To characterize the distributions in Figure~\ref{fig:exdists}, we define azimuthal brightness contrasts as ratios between specific percentiles of the sampled distributions and a fixed reference level.  For CO, the reference level is chosen to capture the mean intensity of the low-brightness web of emission that pervades the disk and extends between bright, but comparatively rare, filamentary structures (see next section).  Contrasts thus measure the enhancement in CO intensity $I_{\rm CO}$ in these bright filamentary features above the reference level ${\langle I_{\rm CO}\rangle}_{\rm ref}$, where we define 
\begin{equation}
C_{\rm CO} \equiv I_{\rm CO} / {\langle I_{\rm CO}\rangle}_{\rm ref}.  
\end{equation}
An obvious alternative to this definition would be to measure the logarithmic width of the distribution. However, our choice makes it possible to probe the distribution's shape using a series of percentile cuts (in our case, corresponding to increasingly higher density thresholds), and is also less sensitive to the inclusion of low signal-to-noise measurements, which might contaminate the observed distribution with noise.  

Our definition is designed to quantify the relative strength of high brightness enhancements above a floor of low-brightness emission in the cold gas and the stellar disk. At these $150$~pc scales, these enhancements may be related to large-scale motions in the galactic potential or other causes, not only turbulence. We deliberately avoid evoking the link between our measured distribution width and turbulent motions in the gas. Turbulence is commonly considered as a main driver of density variations when studying the parsec-scale substructure of gas and gas motions within molecular clouds \citep[e.g.,][]{kainulainen}, but here its dominant role is far less clear.

\subsubsection{The reference level}
\label{sec:referencelevel}
The CO reference ${\langle I_{\rm CO}\rangle}_{\rm ref}$ in a given radial bin is calculated as the mean of all pixels below the 84th percentile of the CO distribution at that radius in order to focus on the low-brightness side and avoid the brightest emission in the maps.  The choice of the 84th percentile as the threshold is somewhat arbitrary, and areas that appear visually `bright' can sometimes contribute to the measured reference level.  However, one advantage of our approach is that the same threshold definition can be applied uniformly to all bins and corresponds to a robust detection of CO emission everywhere across our maps.  

Another advantage is that reference definition can be applied analogously to the 3.6~$\mu$m images, where the  84th percentile captures the mean `interarm' level of the stellar light.  In each radial bin, we calculate the 3.6~$\mu$m reference level ${\langle I_{3.6 \mu {\rm m}}\rangle}_{\rm ref}$ as the mean of all pixels below the 84th percentile of the 3.6~$\mu$m distribution and use this value to compute the 3.6~$\mu$m contrast  
$C_{3.6 \mu {\rm m}} \equiv I_{3.6 \mu {\rm m}} / {\langle I_{3.6 \mu {\rm m}}\rangle}_{\rm ref} $.  

We investigated other definitions of the reference level, such as selecting the distribution median (50th percentile) or altering the choice of threshold. Overall, we found that while modifying the details of the method can change the precise contrast values, it does not qualitatively alter the results. We prefer the use of the mean below a threshold because we found that the median often falls below the noise level of the map. Taking the mean reduces the impact of noise and offers a more robust reference level, but as mentioned the two approaches yield qualitatively similar results.

In calculating the reference level, segments of radial bins that fall outside the ALMA-observed area are ignored. Non-detections (i.e.\ sightlines within the ALMA-observed area but where no significant emission is identified, which are hence blanked in the PHANGS broad maps) are assigned a value of zero. Note, however, that because we use the high completeness ``broad mask'' version of the PHANGS--ALMA moment-0 maps, only regions with no detection at any resolution remain blank. Thus, these tend to be regions where we have high confidence that little or no CO emission is present. In exchange for this high completeness, these ``broad mask'' maps contain more noise than a more restrictive mask, but our analysis method accounts for this.

Although we have confidence in the broad maks, to be conservative we omit the radial bins with a high fraction of non-detections from our analysis. We remove bins that fall below a conservative CO detection fraction threshold of 50\% from the calculations. In total, such bins make up $\sim 25\%$ of the total contrast measurements in our sample and come mainly from a few specific galaxies: NGC~1097, NGC~1365, NGC~1672, NGC~2566, NGC~4536, NGC~4569, and NGC~4731. The radial bins with many non-detections consistently arise near map edges, and their inclusion can in some cases clearly depress the reference level below a reasonable value and push the CO contrasts beyond the vertical range of Figure~\ref{fig:contrasts}.\footnote{In a majority of these annuli, the CO spatial distribution is highly asymmetric, rather than spread evenly throughout the bin, with little in common with the more symmetric stellar morphology.}

For the remaining bins with $>\!50$\% of their pixels included in the moment-0 maps, CO contrast levels are not strongly influenced by non-detections. In these cases, the measurements would be almost identical if we replaced the zeros in the CO maps with the local noise level. On average, such a substitution increases the reference level by only $0.02\pm0.04$ dex ($\sim$4\%). In the future, stacking or similar experiments focused on identifying faint emission in PHANGS--ALMA may yield more insight into these apparently empty regions. For this paper, our methodology and selection of radial bins to study appears robust to the treatment of nondetections.

\begin{figure*}[t]
\begin{center}
\begin{tabular}{c}
\includegraphics[width=0.985\linewidth]{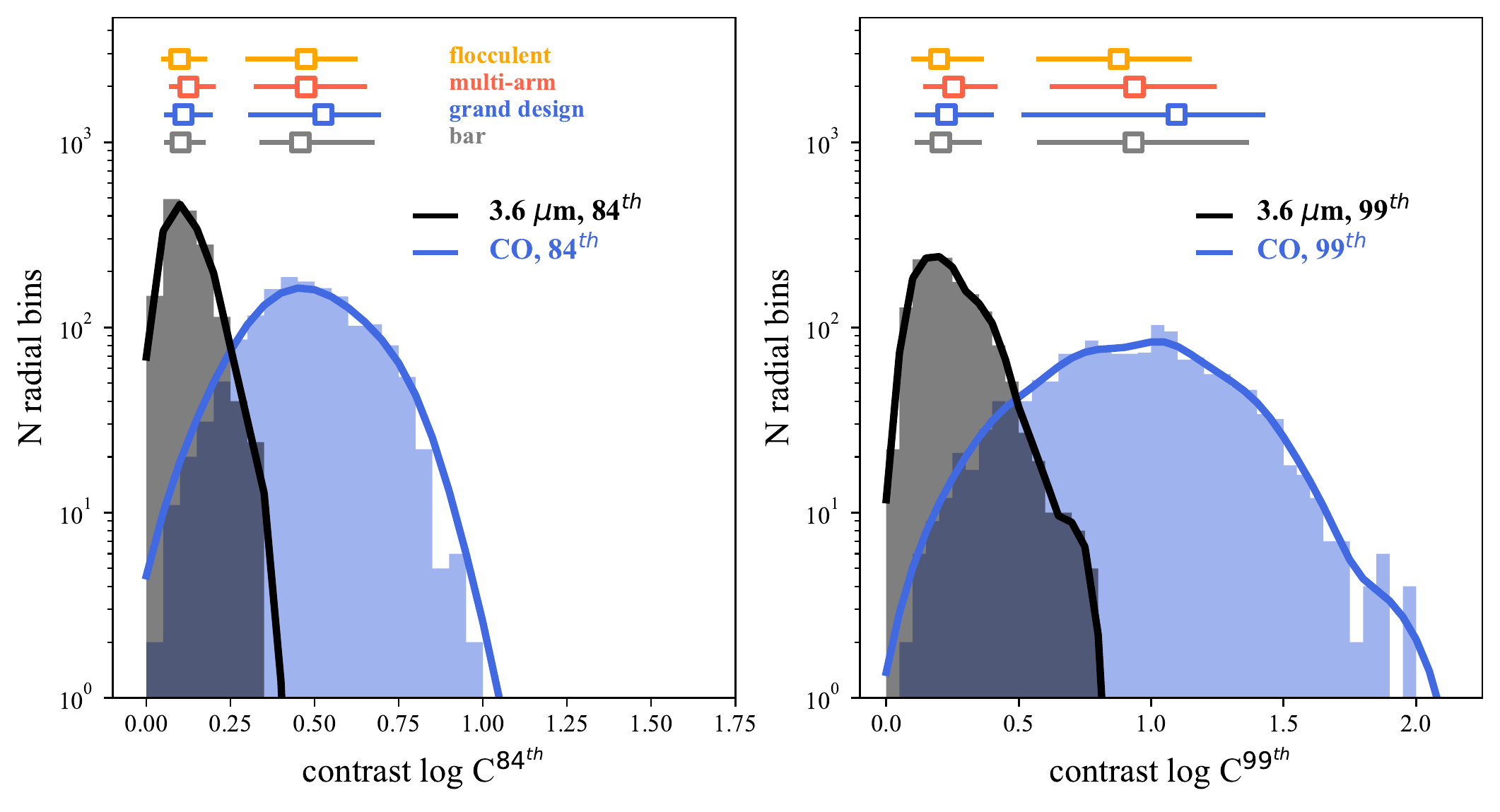}
\end{tabular}
\end{center}
\caption{Histograms of the 150~pc scale $\log C_{\rm CO}$ (blue) and $\log C_{3.6 \mu {\rm m}}$ (black) measured at the 84th (left) and 99th (right) percentiles of the CO and 3.6~$\mu$m distributions sampled in 1527 radial bins across our 67 galaxy sample.  Thick blue and black lines show kernel density estimators of the histograms. The colored symbols and horizontal bars at the top of each panel mark the median and 16th-to-84th percentile width of distributions of contrasts grouped by stellar morphology, and listed in Table~\ref{tab:contrasts}. Rings with CO detection fractions $<\!50\%$ are omitted.}
\label{fig:contrasthisto}
\end{figure*}

\subsubsection{Percentile probes of the distribution}\label{sec:pctlprobes}
For each ring, we measure the contrast between the CO intensity at the 84, 93, 97 and 99th percentile and the reference value.  To obtain an equivalent set of contrasts for the stellar distribution requires an additional modelling step since our nominal stellar tracer is susceptible to contamination from non-stellar emission at the brightness levels probed by our highest percentile cuts. This non-stellar 3.6~$\mu$m emission traces star formation that often originates near the peak in stellar density (as a result of natal gas organization by stellar dynamical features).  Strategies to avoid this kind of contamination often assume that the old stellar light is smoothly distributed (\citealt{elm2011}).  We take a similar approach and estimate corrected percentiles for the 3.6~$\mu$m map by assuming that the true intensity distribution is log-normal, with a centroid and dispersion set by the median and 16th percentile, and then taking the 84, 93, 97 and 99th percentiles from this distribution instead of the observed one (see Figure~\ref{fig:exdists}).  The log-normal shape is chosen on purely empirical grounds, based on inspection of the observed brightness distributions.  This choice allows us to avoid a trivial correlation between CO and 3.6~$\mu$m contrasts due to star formation (see Appendix~\ref{sec:alternatives}), but it may miss further local enhancements in stellar concentration (above the modeled level) possible in spiral arms, for example.  We note here that the contrasts constructed in this way yield qualitatively similar results to those measured from the 3.6~$\mu$m stellar mass maps from \citetalias{Querejeta2015}, although we emphasize that contrasts measured far above the 84th percentile do tend to be sensitive to the treatment of 3.6~$\mu$m dust emission (see \S\ref{sec:results} and Appendix~\ref{sec:alternatives}).  Once the non-stellar 3.6~$\mu$m emission is accounted for, remaining brightness variations predominantly reflect genuine changes in stellar density, as opposed to variations in stellar age, as we argue in Appendix~\ref{sec:agevariations} (see also Appendix~\ref{sec:alternatives}).  

\subsubsection{Relation to morphological contrasts}
\label{sec:contrastcomparison}

The contrasts defined above are measured irrespective of morphological environment, i.e.\ they do not explicitly contrast arm and interarm regions. This approach differs from some previous work aimed at defining, and then contrasting, spiral arm and interarm regions  \citep[e.g.,][]{foyle10,elm2011,Querejeta}. This is a deliberate choice, as we wish to focus this work on quantifying the general divergence or similarity of the CO and $3.6\mu$m morphologies. To this end, a contrast definition that is simple, reproducible, and can be applied to all galaxies is desirable. We have thus chosen a contrast definition that can be applied even in scenarios without traditional spiral structure, and in a way that avoids reliance on environment definitions that are not guaranteed to apply equally well to all galaxies at all wavelengths.

Detailed maps of the morphological features of PHANGS--ALMA targets are presented by \citet{Querejeta} and used to examine the distributions of CO luminosity and gas depletion time in different environments.  \citealt{Sun2020b} and \cite{Rosolowsky2021} explore environmental variations in cloud-scale gas properties using these environment masks. These studies provide a complementary view to our general statistical comparison.

Indeed, we find with testing that contrasts measured in this work yield a picture that is compatible with more traditional `arm-interarm' contrasts defined using the \cite{Querejeta} morphologically-based arm definitions.  These contrasts can be measured for the subset of the galaxies in our sample that host well-defined spiral arms over an extended portion of the area surveyed by PHANGS--ALMA.  For this subset of targets, we calculate the mean CO and 3.6~$\mu$m `arm' and `interarm' brightness levels in each elliptial annulus and measure the contrast as the ratio of the two. In this way we confirm that our calculated reference level performs as intended, falling within $0.15\pm0.18$ dex ($0.05\pm0.1$ dex) on average of the CO (3.6~$\mu$m) `interarm' brightness level. Our measured contrasts exhibit similar levels of agreement with the arm-interarm ratios.

These tests also reveal an expected sensitivity to the arm filling factor in the morphological masks. In the present version of the \cite{Querejeta} masks, the bright CO ridge occupies a lower fraction of the arm area than is filled by the smoother 3.6~$\mu$m emission. With a narrower arm definition, masks unique to each individual tracer would be necessary. In either case, the mean CO ``arm'' intensity depends strongly on the precise arm width and location in the mask.  These arm definitions, it should be noted, are not identical to the masking employed by \cite{foyle10} or \cite{elm2011}.  

Given that we find overall agreement, but a sensitivity of the masking-based approach to the detailed mask definition, we consider the two approaches highly complementary and proceed presenting a more general statistical analysis of azimuthal structure here. We expect future work exploring the links and differences between the two approaches to be fruitful.

\begin{table*}[t]
\begin{center}
\caption{Properties of the measured 150~pc scale CO and 3.6~$\mu$m contrasts}\label{tab:contrasts}
\begin{threeparttable}
\begin{tabular}{rcccc}
\hline
 & 84th percentile & 93rd percentile & 97th percentile & 99th percentile\\
\hline
{\bf $\log C_{\rm CO}$}&&&&\\
\hline
all\tnote{a}&0.5$_{-0.17}^{+0.18}$&0.69$_{-0.24}^{+0.24}$&0.84$_{-0.3}^{+0.29}$&0.98$_{-0.62}^{+0.34}$\\
\hfill bars&0.49$_{-0.15}^{+0.18}$&0.71$_{-0.24}^{+0.27}$&0.89$_{-0.34}^{+0.32}$&1.02$_{-0.42}^{+0.4}$\\
\hfill grand-design&0.53$_{-0.21}^{+0.17}$&0.73$_{-0.34}^{+0.25}$&0.91$_{-0.23}^{+0.3}$&1.1$_{-0.58}^{+0.3}$\\
\hfill multi-arm&0.47$_{-0.15}^{+0.18}$&0.65$_{-0.21}^{+0.21}$&0.8$_{-0.27}^{+0.25}$&0.94$_{-0.3}^{+0.3}$\\
\hfill flocculent &0.47$_{-0.17}^{+0.15}$&0.63$_{-0.23}^{+0.2}$&0.76$_{-0.25}^{+0.23}$&0.88$_{-0.3}^{+0.27}$\\
\hline
{\bf $\log C_{3.6 \mu {\rm m}}$} &&&&\\
\hline
all&0.11$_{-0.05}^{+0.08}$&0.15$_{-0.07}^{+0.10}$&0.19$_{-0.09}^{+0.13}$&0.23$_{-0.11}^{+0.16}$\\
\hfill bars&0.11$_{-0.05}^{+0.08}$&0.15$_{-0.07}^{+0.11}$&0.18$_{-0.09}^{+0.13}$&0.22$_{-0.1}^{+0.16}$\\
\hfill grand-design&0.11$_{-0.05}^{+0.08}$&0.15$_{-0.07}^{+0.11}$&0.19$_{-0.09}^{+0.14}$&0.23$_{-0.12}^{+0.17}$\\
\hfill multi-arm&0.12$_{-0.05}^{+0.07}$&0.17$_{-0.05}^{+0.08}$&0.21$_{-0.09}^{+0.12}$&0.25$_{-0.1}^{+0.16}$\\
\hfill flocculent &0.1$_{-0.04}^{+0.09}$&0.15$_{-0.07}^{+0.11}$&0.19$_{-0.09}^{+0.13}$&0.23$_{-0.11}^{+0.18}$\\
\hline
\end{tabular}
 \begin{tablenotes}
\item[a]Indicated ranges denote the 16th and 84th percentiles of the distribution.  
\end{tablenotes}
\end{threeparttable}
\end{center}
\end{table*}

\begin{figure*}[t]
\begin{center}
\begin{tabular}{c}
\includegraphics[width=1.0\linewidth]{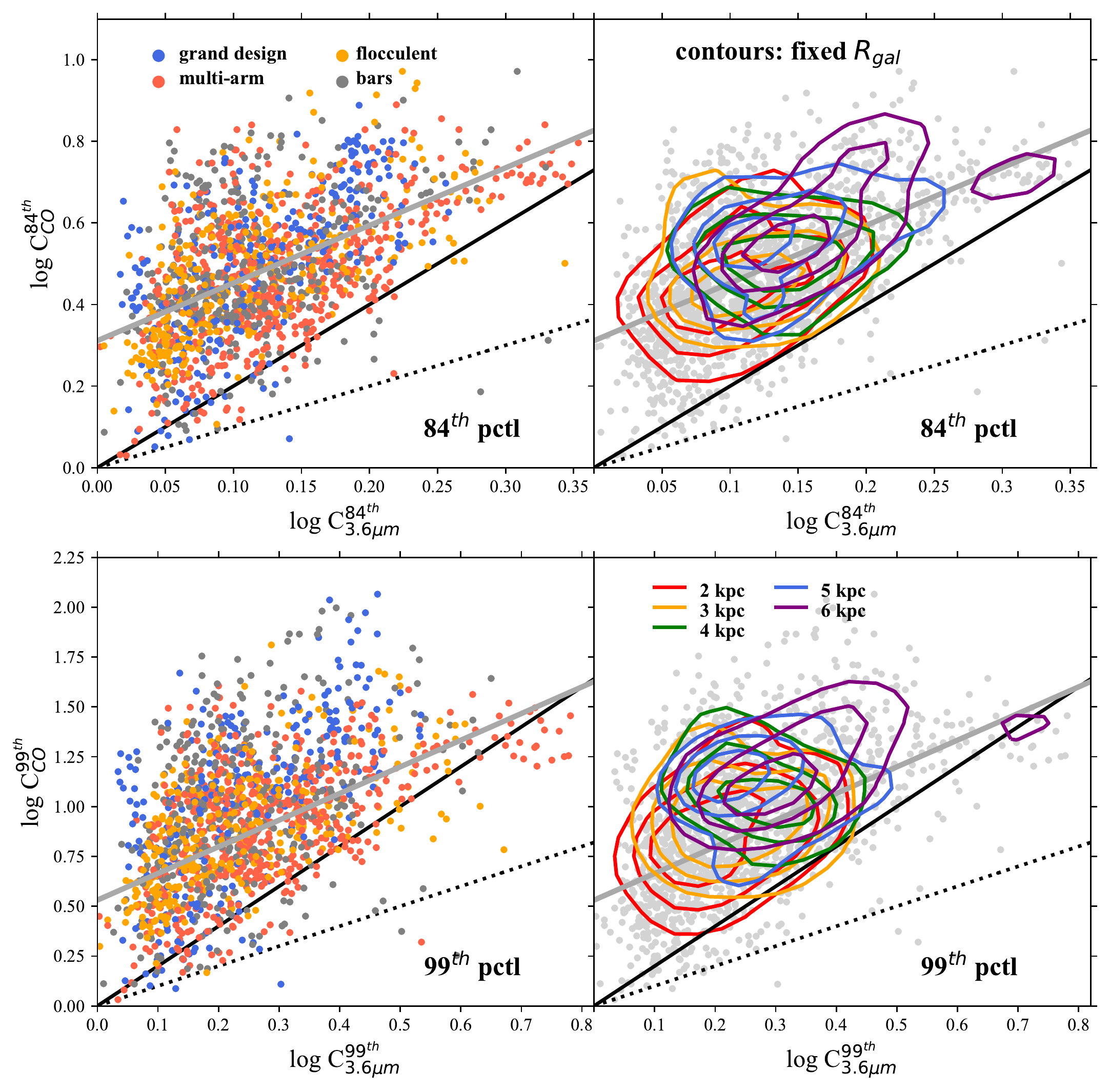}
\end{tabular}
\end{center}
\vspace*{-10px}
\caption{150~pc scale CO contrasts versus baseline 3.6~$\mu$m contrasts measured at the 84th (top) and 99th (bottom) percentiles of each distribution drawn from a set of 150~pc wide elliptical annuli sampling throughout 67 galaxies.  (Left) Points are color-coded by environment. (Right) Five sets of contours highlight the correlation present in the main disk environment at fixed galactocentric radius, spanning in 1~kpc wide rings from 2~kpc to 6~kpc (red to purple).  Rings with detection fractions $<\!50\%$ are omitted from the plot.  
The two black reference lines indicate one-to-one (dotted) and two-to-one (solid) relations in logarithmic space. A third gray reference line shows the best-fit line to all plotted points as given in the text.  Note that the ranges of the axes are different in the top and bottom sets of panels.}
\label{fig:contrasts}
\end{figure*}

\section{Results}\label{sec:results}
\subsection{Molecular gas and stellar density contrasts on the 150~pc scale}\label{sec:general}
In Figure~\ref{fig:contrasthisto}, we plot histograms of the CO and 3.6~$\mu$m contrasts measured from the pixel intensity distributions in 1527 independent radial bins across our sample of 67 galaxies.  These bins are selected to have CO detection fractions $>\!50$\%.  They contain upwards of 1000 pixels each, sampling between 50 and 800 resolution elements per bin.  Contrasts at the 84th and 99th percentiles in each radial bin are shown and summarized in Table~\ref{tab:contrasts}.  The complete list of values for $C_{3.6 \mu {\rm m}}$ and $C_{\rm CO}$ measured using the 84, 93, 97 and 99th percentiles in each radial bin are also presented here as a machine-readable table, of which we show an extract in Table~\ref{tab:mrt} of Appendix~\ref{sec:tabulatedcontrasts}.  

Our set of $C_{3.6 \mu {\rm m}}$ measurements are complementary to the arm-interarm stellar density contrasts measured by \citet{elm2011} and \citet{bittner17}.  To our knowledge, our $C_{\rm CO}$ measurements represent the most extensive set of high resolution molecular gas density contrast measurements for nearby galaxies published to date, surpassing in both number and resolution the arm-interarm contrasts measured from  HERACLES data reported by \cite{foyle10}.     

Figure~\ref{fig:contrasthisto} highlights two notable features of the measured contrasts.  First, the CO contrasts at both percentiles are significantly higher than the corresponding 3.6~$\mu$m contrasts, reproducing the visual impression supplied by the maps (see also Figure~\ref{fig:hists}).  The large average offset between the two sets of contrasts ($\sim$0.5 dex at the 84th percentile, and almost 1~dex at the 99th percentile) appears unlikely to be the result of azimuthal variations in CO excitation.  In normal star-forming galaxies the $\cotwo / \coone$ ratio tends to vary considerably less, closer to the 30\% level \citep{koda20,denBrok2021}, whereas a factor of~2 variation (i.e.\ from arm to interarm) would be needed to reduce the inferred gas contrasts to the level of the 3.6~$\mu$m contrasts.  
Second, we find a significant range of CO contrasts associated with relatively little variation in the 3.6~$\mu$m contrast.  As we will see in the next section, part of the spread in both sets of contrasts is related to mutual systematic variation, raising the possibility that the two may be causally related. 

\subsection{Trends between gas and stellar contrasts on the 150~pc scale }
\label{sec:maintrend}
Figure~\ref{fig:contrasts} plots the CO contrast $C_{\rm CO}$ versus the baseline 3.6~$\mu$m contrast $C_{3.6 \mu {\rm m}}$ in radial bins selected to have CO detection fractions $>\!50$\%. Focussing on this set of annuli with more complete coverage, our goal is to build a representative picture of gas structure in relation to stellar structure.  We solidify this view by combining contrast measurements from all galaxies together, which show how CO contrasts compare with 3.6~$\mu$m contrasts generally.  Later in \S\ref{sec:scatter} we briefly comment on trends internal to individual galaxies.  
  
Two black reference lines trace the power-law relations $C_{\rm CO}\propto C_{3.6 \mu {\rm m}}^n$ with index $n=1$ (dotted) or~2 (solid). Large-scale compression (and ultimately shocking) in response to the passage of density waves, for example, is expected to yield a power-law relation with index $n=2$ \citep[][see \S\ref{sec:discussion}]{ko02}.  A third reference line in each panel traces the best-fit line to all plotted points, given below.  

The formal uncertainties on the contrasts (not shown) are well below the scatter in the measurements, and are dominated by the uncertainty inherent in the sampling of finite numbers of lines-of-sight per bin. We use jackknifing to measure the standard deviation of our contrasts, which are typically $0.05$~dex for contrasts probing the 84th percentile and reach a maximum of ${\sim}0.1$~dex for the 99th percentile.  These uncertainties are comparable to the uncertainty associated with the choice of stellar tracer, as examined in Appendix~\ref{sec:alternatives}.  

The top two panels in Figure~\ref{fig:contrasts} plot the gas and stellar contrasts using color to distinguish different galactic environments. The stellar contrasts probe the underlying large-scale bar/spiral non-axisymmetry at varying strengths.  The measured range in $\log C_{3.6\mu {\rm m}}^{\rm 99th}$ is consistent with the 3.6~$\mu$m arm-interarm contrasts measured by \citet{elm2011} and \citet{bittner17}. Our contrast measurements indicate a less strong
sorting by spiral arm class than found in those previous studies, however.  This may be due to methodological differences, considering that our method probes a range of radii and avoids the bright 3.6~$\mu$m emission that we are less confident in as a stellar mass tracer, given that it can reflect dust emission and age effects.

\begin{figure}[t]
\begin{center}
\begin{tabular}{c}
\includegraphics[width=0.995\linewidth]{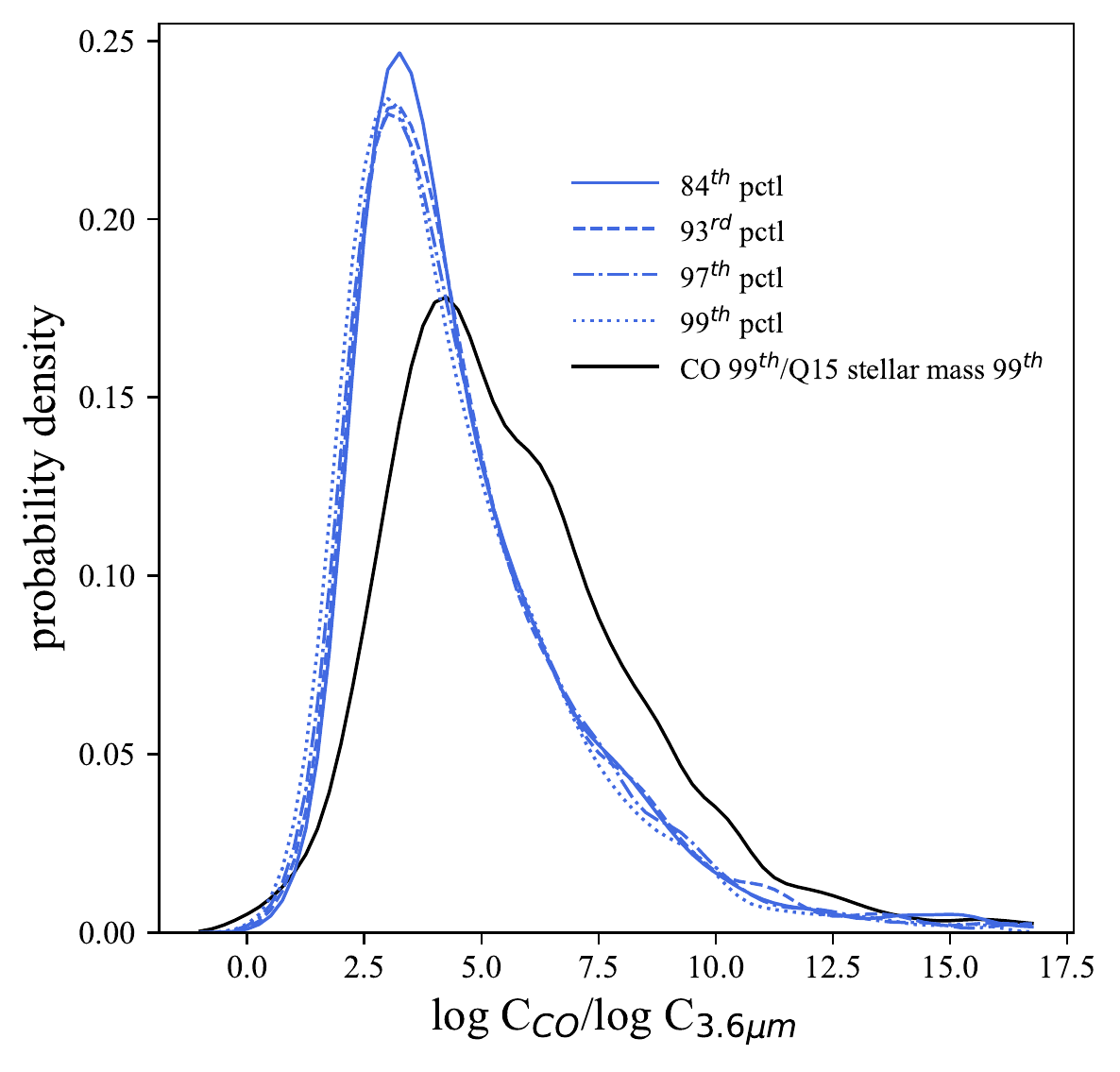}\end{tabular}
\end{center}
\caption{Histograms of the log-contrast ratio $\log C_{\rm CO} / \log C_{3.6 \mu {\rm m}}$ on a fixed 150~pc scale at four CO contrast levels, i.e.\ between the reference level and the 84th (solid blue), 93rd (dashed blue), 97th (dash-dot blue) and 99th (dotted blue) percentiles of the CO and the baseline log-normal 3.6~$\mu$m distributions. An additional reference histogram shows the log-contrast ratio $\log C_{\rm CO} / \log C_{M_\star}$ at the 99th percentile using the \citetalias{Querejeta2015} stellar mass map (solid black).\\} 
\label{fig:hists}
\end{figure}

The CO contrasts, meanwhile, sit well above the matched 3.6~$\mu$m contrasts at each probed percentile (typically 3 to 4 times higher; see Figure~\ref{fig:hists}) and clearly track along with $C_{3.6 \mu {\rm m}}$, though with evident scatter.   
The correlation appears close to ${2\!:\!1}$ on a log--log scale, with a modest vertical offset above the reference ${2\!:\!1}$ line in both panels. 

The correlations at the two percentiles appear equal in strength, as indicated by their similar Pearson~$R$ values $\sim$0.5 that clearly also reflect the scatter evident in the figure (see Table~\ref{tab:pearson}).  The correlation between $C_{\rm CO}$ and $C_{3.6 \mu {\rm m}}$ in all cases is significant, with Pearson~$R$ values more than 10 standard deviations from the null hypothesis constructed by randomly sampling the CO contrast distribution at a given $C_{3.6 \mu {\rm m}}$ ($R=0.0\pm0.02$).  

The qualitative and quantitative similarity of the correlations at different percentiles suggests that gas even at the highest densities is structured by the deep local potential well defined by stellar bars and spirals.  In \S\ref{sec:self-g} we hypothesize that there are two independent aspects of this correlation: (i)~an underlying power-law relation between gas and stellar contrasts and (ii)~scatter that originates with an independent process.  Although we caution that the overall trend would be shaped by these two (and potentially more) processes, we note here the following best-fitting relations (plotted in Figure~\ref{fig:contrasts}; excluding bins with CO detection fractions $<\!50$\%) to the contrasts at the 84th percentile:
\begin{equation}
\log C_{\rm CO}=1.41 \log C_{3.6 \mu {\rm m}}+0.31, \label{eq:84fit}
\end{equation}
with an rms scatter of $\sim$0.15 dex in $C_{CO}$ about the relation,
and at the 99th percentile:
\begin{equation}
\log C_{\rm CO}=1.29 \log C_{3.6 \mu {\rm m}}+0.53, \label{eq:99fit} 
\end{equation}
with $\sim$0.3 dex scatter in $C_{CO}$ about the relation.
Note that these fits do not take into account the possibility of systematic variation in the measurement uncertainties. 

Equations~\eqref{eq:84fit} and~\eqref{eq:99fit} are intended as purely empirical references.  They are not necessarily expected to describe the mechanisms responsible for establishing the trend between CO and stellar contrasts.  Indeed, given the high degree of scatter in the contrasts, the parameters of a power-law relation, or any other hypothesized relation, are not especially well-constrained. We view these fits primarily as useful indicators of the location of the data in contrast-contrast parameter space.

\subsubsection{Sources of scatter}\label{sec:scatter}

The $C_{\rm CO}$ and $C_{3.6 \mu {\rm m}}$ measurements internal to individual galaxies trace out trends that are consistent with, but not necessarily identical to, the overall correlation evident in Figure~\ref{fig:contrasts}.  The weighted average of the slopes of the best-fit lines to individual 84th percentile $\log C_{\rm CO}$ vs.\ $\log C_{3.6 \mu {\rm m}}$ measurements is $1.46\pm0.64$, using the inverse squared errors on the fitted slopes as weights.  The best-fit lines for 99th percentile contrasts are similar.  In all but 26 of the 67 galaxies, fitted slopes differ by less than 3$\sigma$ from the slope of the nominal best-fit line to the full set of contrasts.  Thus, the scatter internal to galaxies, which contributes to diversity in the slopes traced by individual galaxies, is a source of vertical scatter visible in Figure~\ref{fig:contrasts}. This presumably partially reflects the detailed gas response unique to each disk's structure (see Appendix~\ref{sec:gasflowmodel} for a model), yielding gas flows (and local compressions) that vary in strength depending on position relative to the corotation of the underlying disk structure. 

Another part of the scatter in Figure~\ref{fig:contrasts} can be attributed to variation that is present from galaxy to galaxy, according to the diversity in disk structure and gas flows across the sample.  Since we are probing near the molecular cloud scale we can also expect that some of the scatter between otherwise equivalent regions in different galaxies should reflect the finite lifetimes of clouds \citep[i.e.][]{Kruijssen2019b,Chevance2020a}. 
We find that deviation of each galaxy's mean 84th percentile CO contrast from the best-fit relation in Eq.~\eqref{eq:84fit} is comparable on average (0.09 dex) to the average scatter in CO contrast internal to galaxies (0.1 dex).  The two sources of scatter are also comparable ($\sim$0.2 dex) at the 99th percentile.  

In the intra-galaxy variation, we find a weak sorting 
by spiral arm class: at fixed $C_{3.6 \mu {\rm m}}$, CO contrasts are systematically (if only slightly) larger in grand-design spirals than in multi-arm and flocculent spirals.\footnote{Resemblance between the CO distribution and the contribution from dust heated by star formation at 3.6~$\mu$m could suggest a similar sorting may be present in contrasts measured from the observed 3.6~$\mu$m distributions.} This could reflect differences in the magnitude of the local gas flows associated with more tightly-wrapped spiral arms as opposed to open, flocculent spirals (see Appendix~\ref{sec:gasflowmodel}).  It presumably also stems from systematic differences in the strength of gas self-gravitation or other (non-gravitational) forces, which factor into the observed morphological diversity of gas and recent star formation depending on spiral type (e.g.\ \citealt[][]{ko02, dobbspringle, baba15}; see also \citealt{lavigne06,yuho18}).  

Aside from influencing large-scale gas morphology, self-gravity also acts to restructure the gas locally, i.e.\ adding a power-law tail to log-normal gas distributions \citep[]{klessen2000,jaupartchabrier}, such as measured on cloud scales and below \citep{kain09,hughesI,lombardi,schneider}.   We note that 
the scatter and correlation strength change very little at different brightness percentiles, suggesting that the same mechanisms act to organize gas over the large range in density probed here. 

As highlighted in the bottom panels of Figure~\ref{fig:contrasts}, the correlation between $C_{\rm CO}$ and $C_{3.6 \mu {\rm m}}$ is also present at fixed $R_{\rm gal}$, implying that the trend in the top row is not merely the consequence of the mutual radial variation of both quantities.  The scatter about the correlation is moreover roughly fixed at all radii, perhaps indicating that the mechanisms responsible for organizing the gas are active in similar proportions throughout the disk.  

\begin{table*}
\begin{center}
\caption{Properties of the correlation between 150~pc scale CO and 3.6~$\mu$m contrasts}\label{tab:pearson}
\begin{threeparttable}
\begin{tabular}{rcccc}
\hline
 & 84th percentile & 93rd percentile & 97th percentile & 99th percentile\\
\hline
$\langle\Sigma_{\rm H_2}\rangle$~[M$_{\odot}$~pc$^{-2}$]\tnote{a}& 36$_{-24}^{+85}$  & 55$_{-38}^{+135}$  & 78$_{-54}^{+188}$& 107$_{-73}^{+241}$ \\
Pearson~$R$: all&0.54&0.51&0.50&0.50\\
\hfill bars&0.38&0.32&0.36&0.39\\
\hfill grand-design&0.64&0.64&0.63&0.60\\
\hfill multi-arm&0.57&0.56&0.55&0.54\\
\hfill flocculent &0.64&0.60&0.60&0.62\\
\hline
\end{tabular}
 \begin{tablenotes}
\item[a]Adopting a standard Galactic $X_{\rm CO}$ conversion factor \citep{bolatto2013} with $\cotwo / \coone = 0.65$ \citep{denBrok2021}. Indicated ranges denote the 16th and 84th percentiles of the distribution.  
 \end{tablenotes}
\end{threeparttable}

\end{center}
\end{table*}

\subsection{Contrast ratios}
Figure~\ref{fig:hists} shows histograms of the ratio of the logarithmic contrasts plotted in Figure~\ref{fig:contrasts}, i.e.\ $\log C_{\rm CO} / \log C_{3.6 \mu {\rm m}}$.  For reference, Figure~\ref{fig:hists} also shows the histogram of log-contrast ratios with $C_{3.6 \mu {\rm m}}^{\rm 99th}$ replaced by contrasts $C_{M_\star}^{\rm 99th}$ measured from the \citetalias{Querejeta2015} stellar mass maps (where available).  

Overall, at 150~pc resolution, $\log C_{\rm CO}$ sits at a roughly fixed factor of 3 to 4 above $\log C_{3.6 \mu {\rm m}}$ at any of the percentiles probed here.  
The clear correlation in Figure~\ref{fig:contrasts}, which suggests an approximately power-law relation between the CO and 3.6~$\mu$m contrasts, contributes to the narrowness of the log-contrast ratio distributions (with a typical 1$\sigma$ dispersion of 30\%).  Note that both the index $n$ and the normalization $\delta_0$ of a power-law relation $C_{\rm CO}=\delta_0 C_{3.6 \mu {\rm m}}^n$ would factor into the ratio $\log C_{\rm CO} / \log C_{3.6 \mu {\rm m}}$.  

\section{Discussion}\label{sec:discussion}

The distribution of CO brightnesses in the PHANGS-ALMA survey provides a record of the mechanisms that organize the molecular gas on the scales of giant molecular clouds.  Our results suggest that the bulk of the CO-traced molecular gas at 150~pc resolution is organized in relation to the stellar distribution, with a modest restructuring presumably due to self-gravity.  We explore this possibility further in the next two sections, where we develop our findings into a model that relates the observed correlation between CO and 3.6~$\mu$m contrasts to a combination of self-gravity and the galactic potential.  We also discuss a number of other factors that can influence our measured contrasts.  

\subsection{Large-scale gas flows and the prevalence of compression and shocks in molecular gas}\label{sec:shocks}
The clear differences in the measured CO and 3.6~$\mu$m contrasts (see Figure~\ref{fig:contrasthisto}) suggest that the non-axisymmetric stellar dynamical features that lead to the measured 3.6~$\mu$m contrasts do more than sweep up and collect molecular gas.  In such a passive mode, CO brightness contrasts might be expected to be directly proportional to stellar contrasts, as it indeed roughly appears on 500~pc scales \citep{foyle10}.   
However, we have demonstrated that, approaching the cloud scale, CO contrasts are considerably larger than stellar contrasts, perhaps indicating that gas undergoes a much more active mode of organization.  

Compared to preceding generations of CO surveys, PHANGS-ALMA maps probe much nearer to the scales over which gas is compressed on its way to becoming shocked as it flows (supersonically) through stellar bar and spiral structures.  

By probing down to 150~pc scales, we find that CO contrasts are related to 3.6~$\mu$m contrasts in a manner that resembles the relation between large-scale hydrodynamic shock strength and the Mach number of the shock set by the strength of the underlying density perturbation that drives the supersonic flows (e.g.\ \citealt{ko02}; yielding a power-law relation between gas and stellar contrasts with index $n=2$).
Scatter about this trend would naturally arise as a result of self-gravity and magnetic fields, which impact the accumulation of gas at the shock \citep{ko02,baba15}, and presumably feedback as well.  

The pervasiveness of compression and shocking in the cold ISM \citep[e.g.,][]{maclow04} could be responsible for a similar relation generally, even in the absence of density waves.   To lowest order, gas in material spirals moves together with, and supports, the stellar pattern, yielding similar stellar and gaseous overdensities.  But depending on the longevity of the pattern, the gas will shock as it falls towards the local potential minimum established by the stars \citep{baba15}, yielding a shock strength (and gas contrast) proportional to the square of the stellar overdensity.  

We find no considerable difference in the relation between CO and 3.6~$\mu$m contrasts when sorting by morphology or spiral arm class, 
perhaps indicating that stellar overdensities of any origin can similarly compress and focus cold gas.   
Our measurements may thus slightly disfavor especially short-lived material patterns, although we note that other characteristics of shocks, such as length, location, and kinematics \citep{dobbsbaba14,baba15,baba16,pineda} may offer preferable ways to distinguish between density wave and material patterns of varying ages.  

\subsection{The degree of self-gravitation}\label{sec:self-g}
The correlation between CO and 3.6~$\mu$m contrasts found in \S\ref{sec:results} is a clear sign that the host galaxy contributes to the cloud-scale organization of molecular gas.  Such organization moreover indicates that, on the scales of molecular clouds, the host galaxy potential must be at least as important as factors such as self-gravity, the feedback from star formation and other non-gravitational forces.  At our present resolution, these factors appear to introduce the scatter in the observed correlation, which is present even at some of the highest densities probed on the 150 pc scale.  The persistence of the correlation across the range in gas densities studied in this work likely indicates that the present resolution is insufficient to identify exclusively self-gravitating gas.  

In this section we attempt a more quantitative assessment of the strength of the host galaxy potential compared to self-gravity, in particular.  
At the foundation of this exercise is the assumption that the galactic potential establishes a well-defined correlation between CO and 3.6~$\mu$m contrasts and that self-gravity is the only other important factor determining gas surface densities near the cloud scale.  Deviation from the hypothesized correlation then depends on the degree of self-gravitation in the gas.  

\subsubsection{A two-component gas surface density model}\label{sec:twocomponentmodel}
We begin with the expectation that self-gravitating gas on scale $R$ is in a state of rough dynamical equilibrium, with internal motions either reflecting energy equipartition and collapse \citep[e.g.,][]{vs08, ballesteros, ibanez} or resulting from star formation feedback acting to support the gas from within \citep[e.g.,][]{krumholz05,ostrikerShetty}.  In this case, for approximate virial equilibrium, we write  
\begin{equation}
\sigma_{\rm sg}^2=\pi(a_k/5) G R\, \Sigma_{\rm sg}
\end{equation}
with geometric factor 
\begin{equation}
a_k=\frac{(1-k/3)}{(1-2k/5)}
\end{equation}  
following \cite{bertoldimckee}, and assuming a spherical geometry on and below scale~$R$ and an internal gas density profile $\rho\propto R^{-k}$.   Here $\Sigma_{\rm sg}$ and $\sigma_{\rm sg}$ are the surface density and velocity dispersion of the self-gravitating gas.  

Now we consider a scenario in which the gas is structured not only by self-gravity but also an additional factor, namely the background non-axisymmetric galaxy potential.   Assuming the gas once again reaches equilibrium we write 
\begin{equation}
\sigma^2=\pi(a_k/5) G R (\Sigma_{\rm gal}+\Sigma_{\rm sg}).  \label{eq:twocompmodel}
\end{equation}
In this scenario,  
gas motions $\sigma$ on scale~$R$ balance the weight of the gas split into two parts.  One part reflects the gas surface density framed by the galaxy $\Sigma_{\rm gal}$ and the other reflects the gas surface density set up by self-gravity $\Sigma_{\rm sg}$.  In spiral arms, for instance, $\Sigma_{\rm gal}$ represents the level of the filamentary network established via interaction with the underlying stellar spiral and $\Sigma_{\rm sg}$ is the enhancement in surface density above this level due to self-gravity.  

We use the evidence presented in this paper to hypothesize that the surface density $\Sigma_{\rm gal}$ (and the motion in the gas) responds to the underlying stellar potential according to
\begin{equation}
\Sigma_{\rm gal}=\Sigma_{\rm ref}\delta C_{\rm stars}^{\hspace{.05in}n} \label{eq:galconmodel}
\end{equation}
where $\delta$ is a factor close to unity (described more below), $C_{\rm stars}$ is the stellar contrast as defined in the text, $n\approx1-2$ (motivated by Figure~\ref{fig:contrasts}), and $\Sigma_{\rm ref}$ is the mean gas surface density extending between dense filamentary structures equivalent to the `reference level' defined in \S\ref{sec:referencelevel}.  In what follows we will assume that self-gravity is important only above the 84th percentile where it acts to restructure the gas, increasing gas densities by an amount $\Sigma_{\rm sg}$.  

The factor $\delta$ in Eq.~\eqref{eq:galconmodel} in practice allows for some intrinsic offset from, i.e.\ the zero-crossing ${2\!:\!1}$ reference line in Figure~\ref{fig:contrasts} and depends on the details of the gas flow driven in response to the local stellar overdensity (see Appendix~\ref{sec:gasflowmodel} for an example).  

\subsubsection{Gas vs.\ stellar contrasts in the two-component model}
In the two-component model the gas contrast can be written
\begin{eqnarray}
C_{\rm gas} &=&\frac{\Sigma_{\rm gal}+\Sigma_{\rm sg}}{\Sigma_{\rm ref}}\nonumber\\
C_{\rm gas} &=& \delta C_{\rm stars}^n \left( 1+\frac{\Sigma_{\rm sg}}{\Sigma_{\rm gal}} \right) 
\end{eqnarray}
or, in terms of the self-gravitating gas fraction $f_{\rm sg} = \Sigma_{\rm sg}/(\Sigma_{\rm gal}+\Sigma_{\rm sg})$, 
\begin{equation}
C_{\rm gas} = \delta C_{\rm stars}^n \left( \frac{1}{1-f_{\rm sg}} \right). \label{eq:modelcontrast}
\end{equation}

In the context of this two-component equilibrium model, scenarios with $f_{\rm sg} \ll 1$ reflect a dominant galactic component, whereas $f_{\rm sg}$ approaches unity nearer to the condition of complete self-gravitation.  In the latter case, the gas surface density is set predominantly by self-gravity and reflects no organization by the underlying stellar potential.  We denote the intermediate case $f_{\rm sg}=0.5$ as the line of weak self-gravitation.

According to Eq.~\eqref{eq:modelcontrast}, deviation from a reference relation $\log C_{\rm gas} = n\log C_{\rm stars}$ (as in Figure~\ref{fig:contrasts}) would be due to a combination of the factor $\delta$ and the degree of self-gravitation $f_{\rm sg}$. As argued in Appendix~\ref{sec:gasflowmodel}, $\delta$ should be close to unity, and likely a source of scatter about the underlying correlation between gas and stellar contrasts.  The self-gravitating fraction~$f_{\rm sg}$, on the other hand, would likely introduce scatter and an overall vertical shift from the $\log C_{\rm gas} = n\log C_{\rm stars}$ reference line.  In the two-component model, a larger vertical offset reflects higher average $f_{\rm sg}$.  

In reality, feedback and other, non-gravitational factors not incorporated into the model can also lead to departures from the correlation (and from the equilibrium assumed in Eq.~[\ref{eq:twocompmodel}]).  Our goal with the two-component model is to explore the degree of self-gravitation that would be consistent with the observations. 

\subsubsection{Upper limits on the self-gravitating fraction \texorpdfstring{$f_{\rm sg}$}{f\_sg}}\label{sec:modellimitsfsg}
Using Eq.~\eqref{eq:modelcontrast} and the measurements in Figure~\ref{fig:contrasts}, we can place constraints on $f_{\rm sg}$ with a good enough model for~$\delta$, i.e.\ using maximum likelihood techniques to solve for both $\delta$ and $f_{\rm sg}$ that also properly model the uncertainties on both parameters.  Since this is beyond the scope of this work, here instead we will assume that $\delta$ is unity and use the measurements to place upper limits on the average $f_{\rm sg}$ and its variation.  

\begin{figure}[t]
\begin{center}
\begin{tabular}{c}
\hspace*{-.25in}
\includegraphics[width=0.95\linewidth]{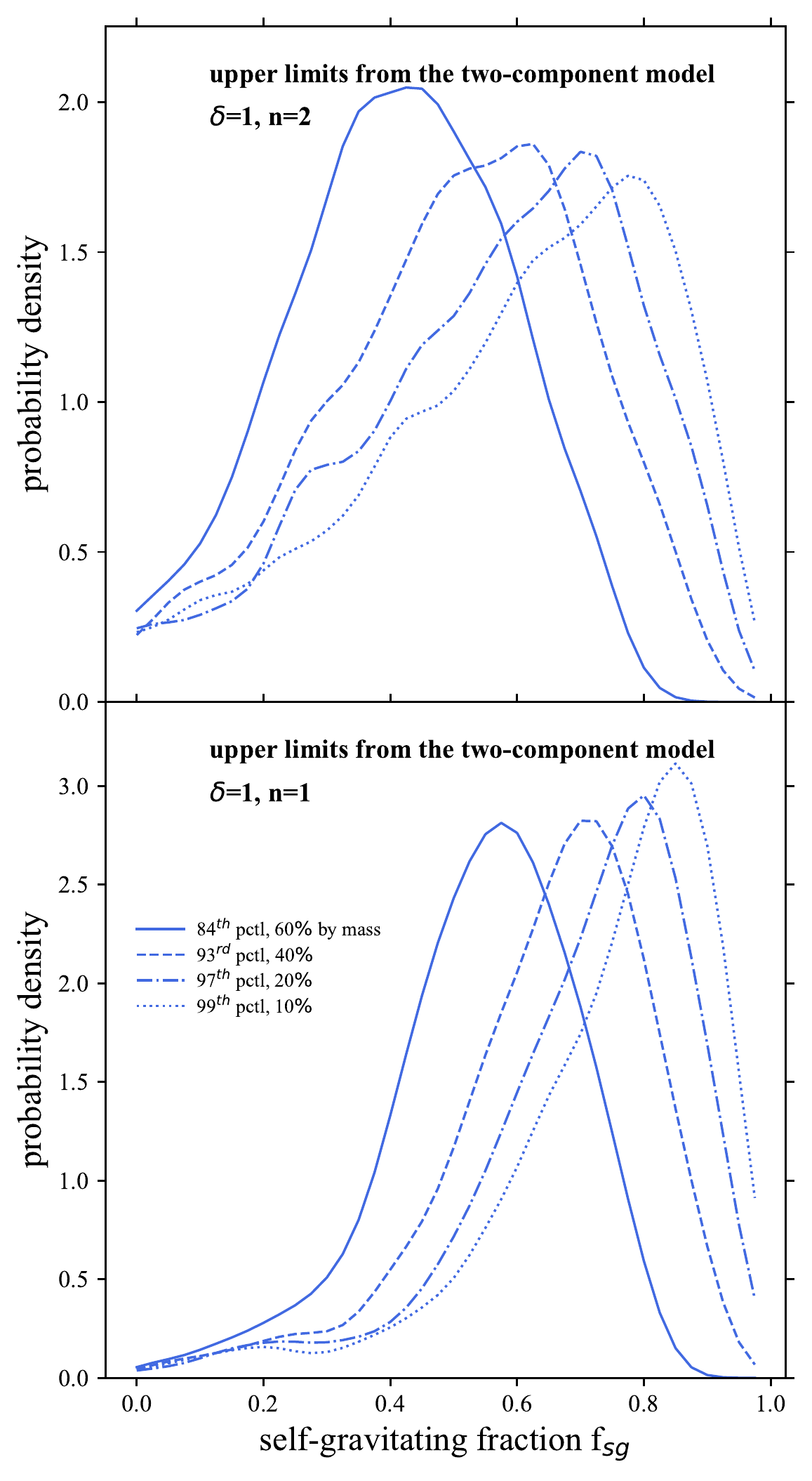}
\end{tabular}
\end{center}
\vspace*{-10px}
\caption{(Top) Histograms of the self-gravitating fraction $f_{\rm sg}$ above four percentiles (84th, 93rd, 97th and 99th) of the CO distribution on 150~pc scales, estimated with the two-component model using the gas and stellar contrast measurements in each of the radial bins plotted in Figure~\ref{fig:contrasts}, assuming $\delta=1$ and active gas organization via background stellar features, with $n=2$.  (Bottom) Same as in the top panel but assuming passive gas organization, with $n=1$. \\}
\label{fig:modeledfsghist}
\end{figure}

Solving Eq.~\eqref{eq:modelcontrast} for the the self-gravitating fraction in terms of the measured $C_{\rm gas}$ and $C_{\rm stars}$ we obtain
\begin{equation}
f_{\rm sg} = 1-\frac{1}{10^{[\log C_{\rm gas}-n \log (\delta C_{\rm stars})]}}.  \label{eq:fsg}
\end{equation}
Histograms of $f_{\rm sg}$ implied by substituting the measurements of $C_{\rm CO}$ and $C_{3.6 \mu {\rm m}}$ in Figure~\ref{fig:contrasts} at either 84th or 99th percentiles for $C_{\rm gas}$ and $C_{\rm stars}$, respectively, are shown in Figure~\ref{fig:modeledfsghist}, where $\delta=1$ is assumed.  Two different scenarios are shown.  The top panel assumes an active mode of gas organization via large-scale disk structure, i.e.\ adopting $n=2$ in Eq.~\eqref{eq:galconmodel}, which denotes the formation of a high density filamentary network during the compression of gas as it flows through passing stellar overdensities, after which contrasts are  elevated to their observed level via self-gravity.  The bottom panel assumes that gas is collected and organized more passively by stellar overdensities, resulting in similar gas and stellar contrasts, i.e.\ 
adopting $n=1$ in Eq.~\eqref{eq:galconmodel}.  In this scenario self-gravity plays a more prominent role in building to the observed CO contrasts.

In the two-component model where $n=2$, our measured CO and 3.6~$\mu$m contrasts imply that the gas at the densities probed by the 84th percentile, which represents $\sim$60\% of the total CO-traced gas mass in the sample, is weakly self-gravitating, with $f_{\rm sg}\approx0.5\pm0.4$ on average.%
\footnote{Note that this is different from the gas being roughly half self-gravitating, which would require half of the radial bins to fall near $f_{\rm sg}=1$.}
The degree of self-gravitation increases at the increasingly higher densities probed by the higher percentiles.  By the 99th percentile, constituting $\sim$10\% of the total CO-traced mass, we find $f_{\rm sg}=0.7\pm0.5$.  

Even in the case that $n=1$, and the estimated degree of self-gravitation is everywhere higher in the gas, retrieving the correlation between $C_{\rm gas}$ and $C_{\rm stars}$ requires that self-gravitation is modest in the bulk of the molecular gas, with $f_{\rm sg} \approx 0.6$ on average above the 84th percentile of the CO distribution.  In this scenario we once again find that $f_{\rm sg}$ increases towards the higher densities probed by the higher percentiles.  

An increase in the degree of self-gravitation towards higher gas densities is also present in the \cite{Sun2020b} accounting of cloud-scale ISM pressure.  We emphasize that the estimates of $f_{\rm sg}$ presented here are meant to offer a rough appraisal of the degree of self-gravitation in the gas that would be consistent with the measured contrasts.   Many other factors not included in the present exercise will most certainly factor into the observed CO contrasts, and the $f_{\rm sg}$ estimates are not expected to be accurate in detail.  

\subsection{Caveats}
In the previous two sections we interpreted the measured CO and 3.6~$\mu$m contrasts as arising from  genuine variations in molecular gas surface density and stellar mass surface density.  In applying both the CO and 3.6~$\mu$m emission as mass tracers, however, we have not formally taken into account the conversion of either to mass.  Indeed, we speculate that variations in either $\alpha_{\rm CO}$ (see \S\ref{sec:general}) or the stellar mass-to-light ratio $\Upsilon_{3.6}$ (see Appendix~\ref{sec:agevariations}) are low enough that they are unlikely to be entirely responsible for the measured contrasts.  However, it should be acknowledged that both factors on the relevant 150 pc scale are undergoing active study, and young stars may locally play a role in setting both (i.e.\ introducing systematic age variations in the stellar population as well as gas temperature changes that alter the $\cotwo / \coone$ ratio and $\alpha_{\rm CO}$).  
We thus tentatively conclude that our interpretation of the measured contrasts as density contrasts is largely valid, but note that this should be reinvestigated in the future.  

\section{Summary \& Conclusion}\label{sec:summary}
We examine contrasts of CO brightness at a fixed 150~pc scale in the PHANGS-ALMA survey in order to probe the factors that influence the distribution of molecular gas surface densities within nearby galaxies.  We match these with `baseline' contrasts in the 3.6~$\mu$m distribution tracing the underlying large-scale stellar dynamical (bar and spiral) features present in these systems.     

We find that the influence of large-scale stellar features is recorded in the distribution of molecular gas surface densities.  These stellar features collect the gas and appear to coordinate large-scale compression (and ultimately shocking) that builds a baseline `high density network' from which the seeds of self-gravitating structures can emerge.  

Our results for molecular gas in the disks of galaxies suggest that a bulk of the molecular mass probed on 150~pc scales is weakly self-gravitating, organized primarily by bars and by spiral arms that behave qualitatively similarly to the models of \citet{dobbsbonnel07}, \citet{smith14,smith20} and \citet{kko2020}. This conclusion is consistent with the observed CO velocity dispersions on this scale, which indicate that gas is in a remarkably uniform but weakly self-gravitating dynamical state \citep{Sun2018,Sun2020a} and shows evidence of dynamical coupling to the host galaxy potential (\citealt{Meidt2018}; and see \citealt{henshaw20}).  

In both spirals and bars, the emergence of regions that sit above the baseline `high density network' may reflect the ability of a small fraction of the gas to become more strongly self-gravitating even on 150~pc scales (with an increase likely at higher densities on smaller scales).  However, a significant portion of the molecular gas in the disks of nearby galaxies appears to be held in a state in which it is unable to collapse and form stars \citep[see also][]{Meidt2020}.  This is consistent with the observation that a significant amount of the CO-traced molecular reservoir does not appear to be actively forming massive stars \citep{schinnerer2019}.  Alternatively, the gas could potentially be on its way to forming stars, as in the framework developed by \cite{Chevance2020a,Chevance2021} (based on cloud lifetimes that are approximately a dynamical time), but this potential star formation is prevented by early feedback.

Our study of contrasts provides another indication that underlying disk structures influence the cloud-scale organization of molecular gas, consistent with the observed sensitivity of cloud properties to galactic environment (\citealt{colombo2014a}; \citealt{Schruba2019}). This influence would also seem to guarantee a strong response to the global galaxy shape and distribution, giving rise to exponential profiles out of otherwise clumpy molecular gas distributions \citep{Leroy2021b}.  Our measurements provide a crucial reference for the new generation of numerical simulations that capture molecular gas and star formation over a wide range of spatial scales.

\bigskip

This work was carried out as part of the PHANGS collaboration.
We would like to thank the anonymous referee for constructive feedback that helped improve the manuscript.
FB and AB acknowledge funding from the European Research Council (ERC) under the European Union’s Horizon 2020 research and innovation programme (grant agreement No.726384/Empire).  RSK and SCOG acknowledge financial support from the German Research Foundation (DFG) via the Collaborative Research Center (SFB 881, Project-ID 138713538) `The Milky Way System' (subprojects A1, B1, B2, and B8). They also acknowledge support from the Heidelberg Cluster of Excellence STRUCTURES in the framework of Germany's Excellence Strategy (grant EXC-2181/1 - 390900948) and from the European Research Council via the ERC Synergy Grant ECOGAL (grant 855130).  JMDK and MC gratefully acknowledge funding from the Deutsche Forschungsgemeinschaft (DFG, German Research Foundation) through an Emmy Noether Research Group (grant number KR4801/1-1) and the DFG Sachbeihilfe (grant number KR4801/2-1). JMDK gratefully acknowledges funding from the European Research Council (ERC) under the European Union's Horizon 2020 research and innovation programme via the ERC Starting Grant MUSTANG (grant agreement number 714907). ES, CF, HAP, TS, DL and TGW acknowledge funding from the European Research Council (ERC) under the European Union’s Horizon 2020 research and innovation programme (grant agreement No. 694343).  
The work of A.K.L. and J.S. is partially supported by the National Science Foundation (NSF) under Grants No.1615105, 1615109, and 1653300, as well as by the National Aeronautics and Space Administration (NASA) under ADAP grants NNX16AF48G and NNX17AF39G.

This paper makes use of the following ALMA data, which have been processed as part of the PHANGS-ALMA \cotwo\ survey: \\
\noindent ADS/JAO.ALMA\#2012.1.00650.S, \linebreak 
ADS/JAO.ALMA\#2013.1.01161.S, \linebreak 
ADS/JAO.ALMA\#2015.1.00925.S, \linebreak 
ADS/JAO.ALMA\#2015.1.00956.S, \linebreak 
ADS/JAO.ALMA\#2017.1.00392.S, \linebreak 
ADS/JAO.ALMA\#2017.1.00886.L, \linebreak 
ADS/JAO.ALMA\#2018.1.01651.S. \linebreak 
ALMA is a partnership of ESO (representing its member states), NSF (USA), and NINS (Japan), together with NRC (Canada), NSC and ASIAA (Taiwan), and KASI (Republic of Korea), in cooperation with the Republic of Chile. The Joint ALMA Observatory is operated by ESO, AUI/NRAO, and NAOJ. The National Radio Astronomy Observatory is a facility of the National Science Foundation operated under cooperative agreement by Associated Universities, Inc.

\appendix
\renewcommand\thetable{A\arabic{table}}
\setcounter{table}{0}   
\section{CO and \texorpdfstring{3.6 \lowercase{$\mu {\rm m}$}}{3.6 micron} contrasts for the full sample of 67 galaxies}\label{sec:tabulatedcontrasts}
In this section we present 150~pc scale CO and 3.6~$\mu$m measurements for the full set of 2192 radial bins sampling throughout our sample of 67 galaxies.  Radial bins are flagged according to CO detection fraction: bins with CO detection fractions $>\!50$\% ($<\!50$\%) are marked with a one (zero). 

\begin{table*}[tb]
\centering
\caption{Tabulated CO and 3.6~$\mu$m contrasts for the full sample of 67 galaxies.}
\begin{tabular}{crccccccccc}
\hline
\hline
Galaxy & R~[arcsec] & \multicolumn{2}{c}{84th percentile} & \multicolumn{2}{c}{93rd percentile} & \multicolumn{2}{c}{97th percentile} & \multicolumn{2}{c}{99th percentile}& flag\\
 & &$\log C_{\rm CO}$ &$\log C_{3.6 \mu {\rm m}}$&$\log C_{\rm CO}$ &$\log C_{3.6 \mu {\rm m}}$&$\log C_{\rm CO}$ &$\log C_{3.6 \mu {\rm m}}$&$\log C_{\rm CO}$ &$\log C_{3.6 \mu {\rm m}}$&\\
 \hline
IC1954 &3.626 &0.27 &0.20 &0.31 &0.28 &0.32& 0.27 &0.37& 0.50 &1\\
IC1954 &6.042 &0.27 &0.10 &0.34 &0.16 &0.41 &0.21 &0.46 &0.22 &1\\
IC1954 &8.46 &0.28 &0.14 &0.32 &0.20 &0.34 &0.24 &0.37 &0.32 &1\\
IC1954 &10.877 &0.36 &0.16& 0.44 &0.21& 0.54 &0.29 &0.59 &0.33 &1\\
\ldots    \\
\hline
\hline
\end{tabular}
\tablecomments{Contrasts are measured as described in \S\ref{sec:thecontrastdef}. Radial bins with CO detection fractions $>\!50$\% ($<\!50$\%) are indicated with a flag of~1~(0).}
\label{tab:mrt}
\end{table*}

\section{The role of age vs. density variations in setting the measured \texorpdfstring{3.6 \lowercase{$\mu {\rm m}$}}{3.6 micron} contrasts}\label{sec:agevariations}
The narrow 3.6~$\mu$m brightness distributions measured throughout our nearby galaxy sample (i.e.\ Figure~\ref{fig:exdists}; neglecting non-stellar emission at 3.6~$\mu$m, as in \S\ref{sec:pctlprobes}) probe modest variations in the underlying old stellar population, which may contain stars of various ages and metallicities.  
Given the relatively small variations in $\Upsilon_{3.6 \mu {\rm m}}$ expected for different stellar ages in this wavelength range \citep[e.g.,][]{belldeJong}, the light at 3.6~$\mu$m shares the NIR's characteristic sensitivity to the more populous old stars than a young component (compared to optical wavelengths).  Below we will examine the implications of this $\Upsilon_{3.6 \mu {\rm m}}$ variation on our measured stellar contrasts.  

We envision a scenario in which the old stellar distribution is uniform, exhibiting no changes with azimuth, and the only change in 3.6~$\mu$m brightness is introduced by a population of 100 Myr old clusters that varies smoothly with azimuth.  If the brightness in any one resolution element is due to one such cluster plus the underlying old stellar light, then we can write the contrast between the main old disk and the local young enhancement as
\begin{equation}
C_{3.6 \mu {\rm m}}=1+({\rm sSFR} \times t_{\rm cl}) \frac{\Upsilon_{\rm 3.6,old}}{\Upsilon_{\rm 3.6,cl}}\label{eq:clustercontrast}
\end{equation}
where $t_{\rm cl} = 100$~Myr is the age of the cluster, $\Upsilon_{\rm 3.6,cl}$ is the mass-to-light ratio appropriate for this age,  $\Upsilon_{\rm 3.6,old}$ is the mass-to-light ratio for the old component and sSFR is the local specific star formation rate associated with the formation of the young cluster.  Taking a reasonable $\Upsilon_{\rm 3.6,old} = 0.5$ and adopting $\Upsilon_{\rm 3.6,cl} = 0.2$ \citep{norris16}, then according to Eq.~\eqref{eq:clustercontrast}, a minimum $\mathrm{sSFR} \approx 10^{-9}~{\rm yr}^{-1}$ would be required to reproduce $\log C_{3.6 \mu {\rm m}} = 0.1$, which is toward the  low end of the measured 3.6~$\mu$m contrasts (see \S\ref{sec:results}).  This sSFR is one to two orders of magnitude larger than what is observed in nearby galaxies (see \citealt{Leroy2019}).  We therefore conclude that the contrasts measured from the 3.6~$\mu$m distributions in our sample are not a consequence of age variations but reflect genuine changes in the underlying stellar density.

\section{Contrasts measured from alternative \texorpdfstring{3.6 \lowercase{$\mu {\rm m}$}}{3.6 micron} distributions}\label{sec:alternatives}
In \S\ref{sec:pctlprobes} we introduced an empirically motivated log-normal model for the distribution of 3.6~$\mu$m emission from the old stars sampled in a given radial bin in order to avoid the emission from dust heated by star formation that contaminates the observed distributions.  This approach attributes any non-log-normal structure present at high brightness in the observed distributions to dust emission.  
However, this neglects any genuine underlying enhancements in brightness that, i.e.\ introduce secondary peaks in the distribution at high brightness.  A more optimal view of the old stellar distribution that would be able to retain such features would require applying a spatially-varying M/L to the observed 3.6~$\mu$m distribution to account for the non-stellar emission (as well as variation in the properties of the underlying stellar population).  The calibration of such an M/L  on the relevant spatial scales is a topic of ongoing work and beyond the scope of the present paper (see \cite{Leroy2021b} for a calibration).  Here we present two alternative views of the stellar distribution and examine the impact these alternatives have on the measured 3.6~$\mu$m contrasts.  

\begin{figure*}[t]
\begin{center}
\begin{tabular}{c}
\includegraphics[width=0.95\linewidth]{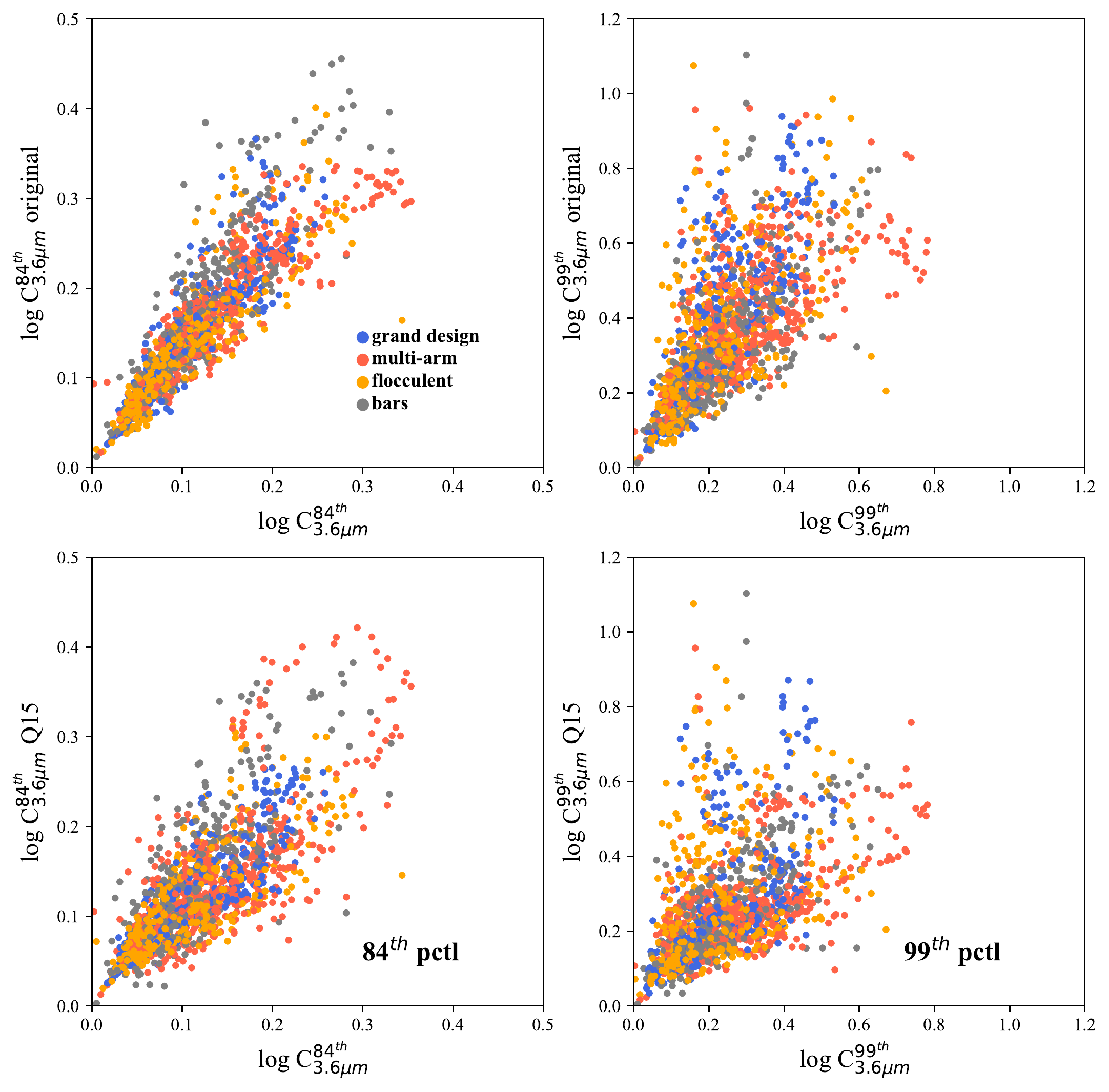}\
\end{tabular}
\end{center}
\caption{Comparisons between our nominal 3.6~$\mu$m contrasts (measured as described in \S\ref{sec:pctlprobes}) and alternative 3.6~$\mu$m contrasts (top: measured from the observed 3.6~$\mu$m distribution; bottom: measured in the \citetalias{Querejeta2015} old stellar light maps, where available) at the 84th (left) and 99th (right) percentiles of each distribution, drawn from a set of 150~pc wide elliptical annuli sampling throughout our sample.  The color coding of points is the same as shown in Figure~\ref{fig:contrasts}.}
\label{fig:altcontrasts}
\end{figure*}

Our first alternative in all sampled radial bins is the observed 3.6~$\mu$m distribution, including any non-stellar emission, which provides an upper limit on the true 3.6~$\mu$m contrast.  The second alternative is supplied by the `old stellar light maps' constructed by \citetalias{Querejeta2015} and introduced in the main text.  These maps are derived with a technique that separates the light in a given sampled pixel into two sources, old stellar light and non-stellar emission, using a combination of two IRAC images (at 3.6 and 4.5 $\mu$m) to determine an optimal global [3.6]--[4.5] color for each of the two sources in all other pixels across the mapped area.  As the global colors are a compromise determined for all pixels, local deviations from these values (i.e.\ due to age and metallicity variations in the stellar population, in the properties of the dust and in the number of independent sources) introduce uncertainty in the mapped distribution of old stellar light (see \citetalias{Querejeta2015}); non-stellar contamination may remain in some regions and be over-subtracted from others.  By contrasting these images with the observed 3.6~$\mu$m maps, though, we gain a useful sense of the global extent of non-stellar contamination.  

The top and bottom panels of Figure~\ref{fig:altcontrasts} show the 3.6~$\mu$m contrasts (from Figure~\ref{fig:contrasts}) versus the 3.6~$\mu$m contrasts measured from our two alternative 3.6~$\mu$m distributions at the 84th (left panel) and 99th (right panel) percentiles.  
Although the \citetalias{Querejeta2015} contrasts better resemble the nominal modeled contrasts than the observed contrasts, the overall agreement in both sets of panels is good.  For contrasts at the 84th (99th) percentile, we measure a 0.05 dex (0.2 dex) standard deviation in the top panel and 0.04 dex (0.16 dex) in the bottom.  

With either of these alternatives, we find no substantial difference in the correlation to CO contrasts highlighted in Figure~\ref{fig:contrasts}.  The only notable discrepancies appear in the observed contrasts above the 99th percentile, at high $C_{3.6 \mu {\rm m}}$.  In some radial bins, enhancements are as large as 0.4 dex and place the measured $C_{3.6 \mu {\rm m}}$ values much closer to the measured $C_{\rm CO}$ values.  This is a sign that the 3.6~$\mu$m brightnesses near the 99th percentile of the 3.6~$\mu$m distribution are contaminated by non-stellar emission, which we can expect to resemble the molecular gas reservoir traced by CO (i.e.\ the gas reservoir ultimately determines the local rate of star formation responsible for heating the dust).  

We find much better consistency among 84th percentile contrasts, and for this reason we emphasize that the stellar contrasts in Figure~\ref{fig:contrasts} are most robust at (and below) the 84th percentile, and sensitive to the treatment of 3.6~$\mu$m dust emission at higher percentiles.  The measurements of the self-gravitating fraction $f_{\rm sg}$ in \S\ref{sec:modellimitsfsg} (and Appendix~\ref{sec:sgmodel}) should thus be treated with caution above the 84th percentile.  
When the full 3.6~$\mu$m emission is sampled (including non-stellar emission),  $f_{\rm sg}$ values estimated above the 93rd, 97th and 99th percentiles are lower than estimated using the \citetalias{Querejeta2015} maps or in Figure~\ref{fig:modeledfsghist}. 

As a third alternative, we have also considered maps of stellar mass surface density constructed via spectral fitting of the MUSE data available for our targets (E.~Emsellem et al., in prep.).  A total of 18 such maps overlap with our \cite{Lang20} parent sample. 
The MUSE mass maps explicitly take into account age variations in the stellar light and are considerably smoother than the original 3.6~$\mu$m maps.  With a few exceptions, these maps reveal bar and subtle spiral arm features that are similar to the structure in the \citetalias{Querejeta2015} maps.  The contrasts we measure from the MUSE mass maps are quantitatively similar to our nominal contrasts.  We find similar levels of variance at the 84th and 99th percentiles as measured for the alternatives shown in Figure~\ref{fig:altcontrasts}.  

All three alternative stellar density tracers considered here show similar degrees of variance from our nominal contrasts, and this variance is comparable to the formal jackknife variance associated with our sampling of the data in narrow radial bins.  We thus consider 0.05 dex and 0.15 dex as good indications of the random uncertainties on our nominal contrasts at the 84th and 99th percentiles, respectively.  

\section{Elements of the two-component gas surface density model}\label{sec:sgmodel}
In this section we discuss in greater detail several of the elements of the two-component gas surface density model introduced in \S\ref{sec:twocomponentmodel}.  

\subsection{Virial equilibrium}
At the basis of the two-component model is the assumption that the molecular gas in galaxies is in virial equilibrium.  Below we discuss why this choice is appropriate even when the gas is not fully self-gravitating and instead also coupled to the galaxy.  

\subsubsection{The gas surface density set up in response to the galactic potential}
The virial equilibrium in Eq.~\eqref{eq:twocompmodel} is separated into two parts.  The first term on the right represents the equilibrium achieved in the presence of gas self-gravity, alone.  The second term $\pi(a_k/5) G R\, \Sigma_{\rm gal}$ represents the equilibrium that can be attained in the presence of the background host galaxy potential.  This term is related to the part of the potential energy that balances the energy in galactic orbital motions (framed by the galaxy potential) in the model of \citet{Meidt2018,Meidt2020}.  As we are interested in scenarios with locally enhanced non-circular orbital motions (due to stellar bar and spiral structures), here in Eq.~\eqref{eq:twocompmodel} we have made the further step that allows the gas to achieve equilibrium in the presence of such motions through an adjustment to its surface density.  This adjustment is the result of the change in shape experienced by clouds that are strongly influenced by the galactic potential.  Clouds are predicted to become triaxial (with semi-minor to semi-major axis ratios $<\!1$ in the plane) given the characteristic shapes of the epicycles traced out by the parcels of gas contained in the cloud as they orbit around the galaxy \citep{Meidt2018}.  
The result is an increase in the surface density in the plane of an initially approximately spherical cloud of fixed mass as it, e.g.\ locally interacts with a spiral arm.  The results of this paper suggest that the gas surface density adjustment obeys Eq.~\eqref{eq:galconmodel}, at least empirically.  

\subsubsection{Relation to self-gravitation}
In the two-component model, gas is fully capable of achieving virial balance on scale~$R$ (as written in Eq.~[\ref{eq:twocompmodel}]) even when it is not fully self-gravitating (i.e.\ when $f_{\rm sg}\neq1$) on this scale.  In this light, observed virial parameters $\alpha = 5\sigma^2 R/(GM)$ in the range of 1 to 2 would not guarantee that gravitational collapse is imminent.  Virial parameters largely in excess of one, meanwhile, would reflect the reality that, since gas is collisional, gas densities are not guaranteed to `adjust' to achieve equilibrium in a given kinematic environment; the coordination of external factors controlling local gas accumulation (such as cloud collisions and agglomerations in spiral arms or turbulent accretion; i.e.\ \citealt{dobbsarms,ibanez}) determine whether equilibrium is reached.  An alternative path for gas with $\alpha>$1 to eventually reach equilibrium involves the evolution of the kinematic environment established by the external gravitational potential (i.e.\ during passage from spiral arm to interarm).  In the present model, such gas would go on to achieve full self-gravitation only when and where (at certain scales and gas densities) the external potential becomes negligible. 

\subsection{The relation between gas contrasts and gas flows}\label{sec:gasflowmodel}
In section \S\ref{sec:shocks} we suggest that the correlation between gas contrasts and stellar contrasts found in this paper arises from the compression and shocking of gas as it flows through passing stellar density features.  In \S\ref{sec:twocomponentmodel} we let the details of the gas flow be encapsulated by the factor $\delta$ in Eq.~\eqref{eq:galconmodel}, and use this parameter in practice to allow for scatter about, and an intrinsic offset from, the zero-crossing ${2\!:\!1}$ reference line in Figure~\ref{fig:contrasts} (assuming $n=2$ in Eq.~[\ref{eq:galconmodel}]).  

In this section we describe the factor $\delta$ in greater detail, by considering the gas flow expected in response to a local stellar overdensity.  In the density wave paradigm, for example, the gas flow at radius~$R$ transverse to a tightly-wound spiral density wave peaking at some angle $\theta_{\rm sp}$ has velocity
\begin{equation}
v_{\rm T} \approx \left[ \sin{i_{\rm p}} V_{\rm c}(R) \left( 1-\frac{\Omega_{\rm p}}{\Omega(R)} \right) \right] (C_{\rm stars}-1) \cos (\theta-\theta_{\rm sp}) \label{eq:dwmodel}
\end{equation}
following \cite{BT}, where $i_{\rm p}$ is the pitch angle of the spiral, $V_{\rm c}(R)$ is the disk rotational velocity at~$R$, $\Omega_{\rm p}$ is the pattern speed of the spiral and $\Omega (R) = V_{\rm c}/R$ is the orbital angular velocity.  As gas flows toward the pattern, a shock forms upstream of the density wave's maximum as the flow becomes transonic, which results in an azimuthal offset between the gas peak $\theta_{\rm gas}$ at the shock and the stellar density peak (tracing the density wave's maximum; i.e.\ \citealt{roberts69,ko02,gittins,baba15}).  This also implies that $v_{\rm T}(\theta_{\rm gas})$ at the site of gas compression will be ${\sim}1{-}4$ times the effective gas sound speed $\sigma$ (i.e.\ Mach numbers $M\approx1{-}4$) in order for the shock to form.  Since the gas density contrast will be proportional to the square of the Mach number $M^2 = (v_{\rm T}/\sigma)^2$ \citep[for radiative shocks; e.g.,][]{ko02}, we expect  $\delta=(\Sigma_{\rm gal}/\Sigma_{\rm ref}) / C_{stars}^n\approx(v_{\rm T}/\sigma)^2/C_{\rm stars}^n\approx 1$ on the way to shock formation in the case that $n=2$, adopting typical values of $C_{star}$ in the range of 1 to 4 (according to $C_{3.6 \mu {\rm m}}$ measurements in \S\ref{sec:results}).

Departures away from $\delta=1$ encode systematic variation in the correlation between $C_{\rm CO}$ and $C_{3.6 \mu {\rm m}}$ with radius, according to the displacement of gas relative to the corotation radius $R_{\rm CR}$, where $\Omega_{\rm p} = \Omega$ (see Eq.~[\ref{eq:dwmodel}]).  Indeed, within individual galaxies a straight $C_{\rm CO}\propto C_{3.6 \mu {\rm m}}^2$ would not be expected.  This will be a source of scatter in Figure~\ref{fig:contrasts}, since the pattern of streaming motions and the location of corotation varies from pattern to pattern and galaxy to galaxy.  It is worth noting that large-scale gas compression as a result of the passage of a density wave should be absent at $R_{\rm CR}$.  Thus, large gas contrasts at this radius, in particular, rely on organization via self-gravity and additional mechanisms that can supply the region with gas \citep[e.g.,][]{beuther,herrera}.

\end{document}